\documentclass{pinchcr}   

\usepackage{graphicx}        
\usepackage{amsmath}

\newcommand{\taua}{\tau_{\alpha}}
\renewcommand{\phi}{\varphi}

\newcommand{\be}{\begin{equation}}
\newcommand{\ee}{\end{equation}}

\title{Glassy dynamics and dynamical heterogeneity in colloids}

\author{Luca Cipelletti}
\affiliation{LCVN, UMR 5587 Universit\'{e} Montpellier 2 and CNRS, P.
Bataillon 34095 Montpellier, France.}

\author{Eric R. Weeks}
\affiliation{Department of Physics,Emory University; Mail stop 1131/002/1AB 400 Dowman Dr.  Atlanta GA 30322-2430 USA.}

\begin{document}

\maketitle

\preface
Concentrated colloidal suspensions are a well-tested model system
which has a glass transition.  Colloids are  suspensions of small
solid particles in a liquid, and exhibit glassy behavior when the
particle concentration is high; the particles are roughly analogous
to individual molecules in a traditional glass.  Because the
particle size can be large (100 nm - 1000 nm), these samples can be
studied with a variety of optical techniques including microscopy
and dynamic light scattering.  Here we review the phenomena
associated with the colloidal glass transition, and in particular
discuss observations of spatial and temporally heterogeneous
dynamics within colloidal samples near the glass transition.
Although this Chapter focuses primarily on results from
hard-sphere-like colloidal particles, we also discuss other
colloidal systems with attractive or soft repulsive interactions.

\section{Colloidal hard spheres as a model system for the glass transition}

\subsection{The hard sphere colloidal glass transition}
\label{colloidalglass}

When some materials are rapidly cooled, they form an amorphous solid
known as a glass.  This transition to a disordered solid is the
glass transition \shortcite{gotze92,stillinger95,ediger96,angell00}.  As
the temperature of a molecular glass-forming material is decreased
the viscosity rises smoothly but rapidly, with little apparent
change in the microscopic structure \shortcite{ernst91,vanblaaderen95}.
Glass formation may result from dense regions of well-packed
molecules or a decreasing probability of finding mobile regions.
As no structural mechanisms for this transition have been found,
many explanations rely on dynamic mechanisms.  Some theoretical
explanations focus on the idea of dynamical heterogeneities
\shortcite{gotze92,sillescu99,ediger00,gibbs65}.  The underlying
concept is that, for any molecule to move, all molecules within a
surrounding region must ``cooperate'' in their movement.  As the
glass transition is approached the sizes of these regions grow,
causing the rise in macroscopic viscosity \shortcite{gibbs65}.  The
microscopic length scale characterizing the size of these regions
could potentially diverge, helping explain the macroscopic viscosity
divergence.  However, it is also possible that these regions could
grow but not be directly connected to the viscosity divergence.
Additionally, it is not completely clear if the viscosity itself
diverges or simply becomes too large to measure \shortcite{hecksher08}.
While the existence of dynamical heterogeneities in glassy
systems has been confirmed in a wide variety of systems,
the details of this conceptual picture remain in debate
\shortcite{sillescu99,ediger00,glotzer00,ngai99,richert02,CipellettiJPCM2005}.

Colloidal suspensions are composed of microscopic-sized solid
particles in a liquid, and are a useful model system for studying
the glass transition.  In terms of interparticle interaction,
the simplest colloids are those in which the particles interact as
hard spheres, i.e., the interparticle potential arises solely due
to excluded volume effects \shortcite{pusey86}. Hard spheres are a
useful theoretical model for glass-forming systems due to their
simplicity \shortcite{bernal64}.  Clearly, attractive interactions
between atoms and molecules are responsible for dense phases of
matter.  But given dense states of matter, repulsive interactions
play the dominant role in determining the structure.  Hard spheres
are useful simulation models for crystals, liquids, and glasses, although it is still debated whether a purely repulsive interparticle potential is sufficient to reproduce the glass transition in general~\cite{berthier09}.

The control parameter for hard sphere systems is the concentration,
expressed as the fraction, $\phi$, of the sample volume occupied by
the particles. Most colloidal hard sphere systems act like a glass
for $\phi$ larger than an ``operational'' glass transition volume
fraction $\phi_g \approx 0.58$. The transition is the point where
particles no longer diffuse through the sample on experimentally
accessible time scales; for $\phi < \phi_g$ spheres do diffuse at
long times, although the asymptotic diffusion coefficient $D_\infty$
decreases sharply as the concentration increases
\shortcite{bartsch93,vanmegen98,bartsch98}. The transition at $\phi_g$
occurs even though the spheres are not completely packed together;
in fact, the density must be increased to $\phi_{\rm RCP} \approx
0.64$ for ``random-close-packed'' spheres
\shortcite{ohern03,bernal64,torquato00,torquato04,ohern04} before the
spheres are motionless. In simulations, a collection of same-sized
(i.e. monodisperse) hard spheres almost always crystallizes. A
binary mixture of spheres or indeed a distribution of sizes is
therefore needed to frustrate crystallization and enable access to
the glass transition, both numerically and
experimentally~\shortcite{henderson98,ZaccarelliPRL2009}.

The glass transition for suspensions of nearly-hard-sphere
colloids
\shortcite{pusey86,vanmegen91,vanblaaderen95,vanmegen93,bartsch93,bartsch95,mason95glass}
is comparable in many respects to the
hard sphere glass transition studied in simulations and theory
\shortcite{speedy98}.
Macroscopically, a colloidal
liquid flows like a viscous fluid whereas the colloidal glass does
not flow easily, like a paste \shortcite{segre95,cheng02}.  For colloidal
samples with low polydispersity, samples can crystallize for
$\phi<\phi_g$ (with crystallization nucleating in the interior of
the sample), while for $\phi>\phi_g$, crystals only nucleate at flat
sample boundaries such as the walls of the container \shortcite{pusey86}.
Microscopically, the glass transition point is identified as the
point where $D_\infty \rightarrow 0$
\shortcite{bartsch93,vanmegen98,bartsch98}.

Note that there are some questions about the colloidal glass
transition.  First, prior measurements disagree about the nature of
the viscosity divergence in colloidal glasses
\shortcite{segre95,cheng02}.  This may be due to difficulties in
reconciling measurements of the volume fraction $\phi$
\shortcite{segre95comment,segre95reply}.  Second, it has been seen that a
glassy colloidal suspension can crystallize under microgravity
conditions \shortcite{zhu97}, and potentially also when in
density-matching solvents \shortcite{kegel00lang,kegel04}, suggesting
that the apparent glass transition at $\phi_g=0.58$ is an artifact
of gravity.  The interpretation of these observations is unclear.
One possibility is that this is merely heterogeneous nucleation at
the walls, as seen before \shortcite{pusey86}.  Another possibility is
that these samples were more monodisperse than most, and a slightly larger
polydispersity ($> 5\%$) may be necessary to induce a glass
transition at $\phi_g=0.58$
\shortcite{henderson98,meller92,auer01b,schope07}. In practice, most
experimental samples always have a polydispersity of at least 5\%.
Recent simulations suggest that the relationship between
polydispersity, crystallization, and glassy dynamics is more complex
than perhaps was previously appreciated \shortcite{ZaccarelliPRL2009}.
Overall we note that the interpretations of the various observations
described in this paragraph are often still controversial.

Colloidal glasses are quite similar to molecular glasses:
\begin{list}{$\bullet$}{\setlength{\itemsep}{0ex}}
\item Both are microscopically disordered \shortcite{vanblaaderen95}.
\item Both are macroscopically extremely viscous near their
      transition point \shortcite{segre95,cheng02}.
\item Colloidal glasses have a nonzero elastic
      modulus at zero frequency,
      which is absent in the liquid phase \shortcite{mason95glass}.
\item Colloidal glasses are out of equilibrium and show aging
      behavior -- their properties depend on the time elapsed since preparation
      \shortcite{courtland03,cianci06ssc,cianci07,kegel04}, similar to
      polymer glasses \shortcite{hodge95,mckenna03} and other molecular
      glasses \shortcite{angell00,castillo07}.
\item Colloidal glasses exhibit dynamical heterogeneity
      \shortcite{kegel00,weeks00}, similar to that seen in simulations
      \shortcite{donati98,glotzer00,doliwa98} and experiments
      on molecular glasses
      \shortcite{ediger00,sillescu99,israeloff00}.
\item The colloidal glass transition is sensitive to finite size
      effects \shortcite{nugent07prl}, similar to molecular glasses
      \shortcite{mckenna05} and polymers \shortcite{roth05}.

\end{list}
One difference between colloids and molecules is that colloidal
particles move via Brownian motion whereas the latter move
ballistically at very short time scales; several simulations
indicate that this difference is unimportant for the long-time
dynamics which is of the most interest
\shortcite{binder98,szamel04,franosch08}.  Likewise, hydrodynamic
interactions between particles influence their motion on short time
scales, but do not modify the pairwise interaction potential (which
remains hard-sphere-like), suggesting that they should not be relevant
for the long-time dynamics~\shortcite{brambilla09}.

One advantage colloids have over traditional molecular
glassformers is that their time scales are significantly slower,
with relaxation taking $O$(1-1000~s), allowing easy study of
the relaxation processes.  A second
advantage of colloids is that their large size [$O$(1 $\mu$m)]
allows for measurement using optical microscopy or dynamic light
scattering, as will be discussed in Sec.~\ref{sec:emethods}.

Experiments have one chief advantage over simulations, in that they
more easily avoid finite size effects.  Near the glass transition,
dynamical length scales can be large ($\sim 4-5$ particle diameters
\shortcite{weeks07cor,doliwa00}) and finite size effects on structure
and dynamics may extend to even larger length scales ($\sim 20$
particle diameters \shortcite{nugent07prl}).  Microscope sample chambers
typically contain $\sim 10^9$ particles, and light scattering
cuvettes contain even more.

The most popular ``hard sphere'' colloid is colloidal PMMA
(poly-methyl-methacrylate), sterically stabilized to minimize
inter-particle attraction \shortcite{pusey86,weeks00,dinsmore01,antl86}.
The particles can be placed in density-matching solvents to inhibit
sedimentation (see Sec.~\ref{challenges}), and/or solvents which
match their index of refraction to enable
microscopy~\shortcite{dinsmore01} and light scattering (see
Secs.~\ref{sec:microscopy} and~\ref{sec:DLS}). In some of these
solvents, the particles pick up a slight charge and thus have a
slightly soft repulsive interaction in addition to the hard-sphere
core. Despite this charge the spheres behave similarly to hard
spheres with some phase transitions shifted to slightly lower $\phi$
\shortcite{gasser01}.  Salt can be added to the samples to screen the
charges and shift the interaction back to more hard-sphere like
\shortcite{yethiraj03,royall03}.

A second popular colloid is colloidal silica, which is relatively
easy to fabricate \shortcite{stober68}.  These particles are suspended in
water (or a mixture of water and glycerol), avoiding the
organic solvents that are required for PMMA colloids.  Because they
are in water, repulsion due to charge is the primary mechanism
preventing flocculation; adding salt can screen the charges and
cause flocculation (which is often irreversible). Silica colloids
are hard to density match (their density varies but is larger than 2
g/cm$^3$), and also it is hard to match their refractive index. This
latter constraint makes microscopy difficult except at lower volume
fractions \shortcite{mohraz08}.

Several glass transition theories have been applied to the colloidal
glass transition.  The colloidal glass transition appears to
be well-described by mode-coupling theory up to $\phi \approx
0.58$~\shortcite{vanmegen91,vanmegen93,vanmegen94,gotze91,schweizer03,schweizer04,saltzman06,saltzman06b}, although other glass transition
theories successfully capture many features of the colloidal glass
transition as well \shortcite{ngai98,oppenheim96,tokuyama95,tokuyama07}.
(See the discussion in Sec.~\ref{sec:relax} which discusses the
strengths and weaknesses of mode-coupling theory as applied to
colloids.)

\subsection{Experimental challenges in studying colloidal hard spheres}
\label{challenges}
As argued in the previous section, colloidal hard spheres are a good
model system for investigating the glass transition. However,
several experimental challenges have to be faced. Probably, the most
serious problem is that of a precise determination of the volume
fraction. In optical microscopy, $\phi$ can be obtained by counting
the number of particles in a given volume and using the particle
size as obtained, e.g., from electron microscopy. It should be noted
that an error of just 1\% in the radius $a$ of the particles results
in a 3\% error in $\phi$. Additionally, one has to take into account
the thickness of the stabilizing layer (e.g. the grafted polymer for
PMMA particles or the counterion cloud for silica particles in a
polar solvent), which is often difficult to measure precisely. Since
typical values of the thickness are on the order of 10
nm~\shortcite{puseyleshouches1991}, this contribution can be relevant for
the small particles used in light scattering ($a \sim 100 - 300$
nm), while it is less important for the micron-sized particles used
in microscopy. Other methods to determine $\phi$ include precise
density or refractive index measurements~\shortcite{chaikin96}. For PMMA,
these methods require special care, since the particles can be
swollen by organic solvents such as tetralin or brominated solvents, which
change their density and refractive index compared to those of the
bulk material.

Because of these difficulties, the absolute volume fraction is often
determined indirectly by comparing the phase behavior or the
dynamical behavior of the sample to theoretical and numerical
predictions. Samples sufficiently monodisperse ($\sigma =
\sqrt{\overline{a^{2}}- \overline{a}^{\,2}}/\overline{a}^{\,2} < 5 -
8\%$) crystallize for $0.494 < \phi <
0.545$~\shortcite{puseyleshouches1991}. The absolute volume fraction can
then be calibrated by matching the experimentally determined
freezing volume fraction with $\phi_f = 0.494$ as determined by
simulations~\shortcite{HooverJChemPhys1968}. However, some uncertainty is
still left, because the exact value of the freezing fraction depends
on
$\sigma$~\shortcite{bolhuis96,fasolo04,vanmegen94,ZaccarelliPRL2009,pusey09}.
Moreover, this method cannot be applied to more polydisperse
suspensions that do not crystallize over several months or years.
Alternatively, $\phi$ may be calibrated against predictions for the
volume fraction dependence of the low shear
viscosity~\shortcite{puseyleshouches1991,poon96} or the short time self
diffusion coefficient~\shortcite{BeenakkerPhysicaA1983,TokuyamaPRE1994}
in the dilute regime. For samples where both the calibration against
$\phi_f$ and that using the short time self diffusion coefficient
are possible, the two methods appear to be
consistent~\shortcite{vanmegen89,SegrePRE1995}. In summary, while
relative values of $\phi$ can be measured very precisely (down to
$10^{-4}$ using an analytical balance), absolute values are
typically affected by an uncertainty of a few \%. This should always
be kept in mind when comparing sets of data obtained in different
experiments.

For colloids, the equivalent of $T = 0$ in a molecular system is
random close packing, the volume fraction $\phi_{\mathrm{RCP}} \sim
0.64$ where osmotic pressure diverges and all motion ceases because
no free volume is left. Clearly, knowledge of the precise location
of $\phi_{\mathrm{RCP}}$ is very important to discriminate between
theories that predict a glass transition below random close packing
(for example the mode coupling theory~\shortcite{GotzeJPCM99} or
thermodynamic glass transition
theories~\shortcite{CardenasJChemPhys1999,ParisiJChemPhys2005}) and
scenarios like jamming, where no arrest is predicted below
$\phi_{\mathrm{RCP}}$. However, the location of random close packing
is still highly
debated~\shortcite{BerthierPRE2009,XuPRL2009,KamienPRL2007,chaudhuri10} and its very
existence is challenged, based on the argument that one can always
trade order for packing efficiency~\shortcite{DonevJChemPhys2007}, up to
$\phi = \pi/\sqrt{18} \approx 0.7405$, the packing fraction of a (monodisperse) hard sphere
crystal. Experimentally, it is difficult to measure
$\phi_{\mathrm{RCP}}$ because of the uncertainty on absolute volume
fractions discussed above and because applying the high pressure
(e.g. by centrifugation) needed to approach it may result in the
compression of the stabilizing layer. Additionally, both
experiments~\shortcite{brujic09} and simulations~\shortcite{schartl94,BerthierPRE2009}
show that $\phi_{\mathrm{RCP}}$ depends on $\sigma$.

Some of the experimental challenges posed by hard spheres stem from
the very same features that make them a valuable model system: their
relatively large time and length scales. The microscopic time in a
colloidal system is the Brownian time $\tau_B$, i.e. the
time required by a particle to diffuse over its own size in a
diluted system, defined as
\begin{equation}
\label{taub}
\tau_B \equiv a^2 / 6 D = \pi \eta a^3 / k_B T \,.
\end{equation} Here, $\eta$ is the solvent viscosity,
$T$ the absolute temperature, $k_B$ Boltzman's constant and $D$ is the diffusion coefficient for a sphere of radius $a$, given by the Stokes-Einstein-Sutherland formula~\shortcite{einstein1905a,sutherland1905}:
\begin{equation}
\label{ses}
D = k_B T / 6 \pi \eta a \,.
\end{equation}
The time scale $\tau_B$ is typically on the order of $10^{-3} - 1$
s. The largest relaxation time that can be measured in concentrated
systems is on the order of $10^5$ s; thus, the accessible dynamical
range covers at most 8 decades, as opposed to 15 decades in
molecular glass formers. Additionally, owing to their relatively
large size and due to the mismatch between their density and that of
the solvent in which they are suspended, colloidal particles
experience gravitational forces that can modify their phase
behavior~\shortcite{zhu97,pusey09} and dynamical
properties~\shortcite{ElMasriJSTAT2009,kegel04}. The relevant parameter
to gauge the importance of sedimentation is the inverse P\'{e}clet
number, $Pe^{-1} = \frac{3k_{B}T }{4\Delta \rho g a^{4}}$, defined
as the ratio of the gravitational length to the particle radius ($g$
is the acceleration of gravity and $\Delta \rho$ the density
mismatch). For diluted suspensions, gravity becomes relevant as
$Pe^{-1}$ approaches (from above) unity; for concentrated
suspensions, sedimentation effects may set in at even higher values
of $Pe^{-1}$, since gravitational stress is transmitted over
increasingly larger distances as the sample becomes more solid-like.
For example, sedimentation effects have been reported to alter the
dynamics of PMMA particles in organic solvents ($\Delta \rho \sim
0.3 \mathrm{g}/\mathrm{cm}^{3}$) for $Pe^{-1} = 44.1$~\shortcite{kegel04}
or even for $Pe^{-1} = 1350$~\shortcite{ElMasriJSTAT2009}. Density-
matching solvents can mitigate these effects, although matching
closely both the index of refraction (as required for optical
observations) and the density of the particles without altering
their hard sphere behavior has been proved difficult for PMMA
particles~\shortcite{royall03} and impossible for other systems such as
silica spheres.

\section{Experimental methods for measuring both the average dynamics and dynamical heterogeneity}
\label{sec:emethods}

\subsection{Main features of optical microscopy and dynamic light scattering}
\label{sec:mic_DLS} Optical microscopy and dynamic light scattering
are the main techniques to probe both the average dynamics and its
spatiotemporal fluctuations in dense colloidal suspensions. Each of
them comes with specific advantages and limitations. Optical
microscopy is unsurpassed in providing detailed information on the
structure and the dynamics at the single particle level. The same
quantities introduced in theory and simulations to characterize the
dynamics can be precisely measured (e.g. the mean square
displacement or the intermediate scattering function for the average
dynamics, and the dynamical susceptibility, $\chi_4$, and the
spatial correlation of the dynamics, $g_4$, for its fluctuations,
see Chapter 2 and Sec.~\ref{sec:cI} below). Additionally, direct
visualization of the sample allows any experimental problem to be
readily detected, such as particle aggregation, sedimentation, or
wall effects. Finally, techniques such as optical or magnetic
tweezing~\shortcite{GrierNature2003,AmblardRevSciInstrum1996} allow one
to manipulate single particles and thus to measure the microscopic
response of the system to a local perturbation~\shortcite{habdas04}.

Dynamic light scattering probes a very large number of particles
simultaneously, yielding very good averages. Moreover, particles
used in light scattering are usually smaller than those for optical
microscopy ($a = 100-500$ nm as opposed to $ a = 0.5-1.5
\mu\mathrm{m}$), which has a twofold advantage. First, the
microscopic time $\tau_B$ (see Eq.~(\ref{taub})) is significantly
reduced, since $\tau_B\sim a^{2}/D \sim a^{3}$, thus increasing
substantially the experimentally accessible dynamical range. Second,
gravitational effects are much less of concern, since $Pe^{-1} \sim
a^{-4}$. Finally, as we will discuss it in Sec.~\ref{sec:cI}, recent
developments allow dynamical heterogeneity to be probed by dynamic
light scattering, although not at the level of microscopic detail
afforded by optical microscopy. These methods extend the
possibilities of light scattering by adding features characteristic
of imaging techniques. Quite in a symmetric way, very recent
microscopy methods such as dynamic differential
microscopy~\shortcite{cerbino08} have extended imaging techniques by
adding the capability of measuring the intermediate scattering
function $f(q,\tau)$ defined below in Sec.~\ref{sec:DLS}.

\subsection{Optical and confocal microscopy}
\label{sec:microscopy}

Given the large size of many colloidal systems (particle radius
$a \sim 100 - 1000$~nm), optical microscopy is a useful tool for
observing these systems.  First, these sizes are comparable to the
wavelength of light, thus rendering them visible.  Second, the time
scales of their motion are often slow enough for video cameras to
follow their motion.  This can be seen by considering how quickly
particles diffuse.  Colloidal particles undergo Brownian motion due
to thermal fluctuations, as discussed in Sec.~\ref{challenges}.
The diffusion coefficent $D$ (Eq.~\ref{ses}) is related to
the mean square displacement of particles as:
\begin{equation}
\label{diffusion}
\langle \Delta x^2 \rangle = 2 D \Delta t,
\end{equation}
where $\Delta t$ is the time scale over which the displacements are
taken. A particle of diameter $2a= 1$~$\mu$m in water ($\eta =
1$~mPa$\cdot$s) at room temperature diffuses approximately 1~$\mu$m
in 1~s. This motion is easy to study with a conventional video
camera and a microscope; most video cameras take data at 30 images
per second, thus they allow one to follow the Brownian motion of
colloidal particles of these sizes.  Of course, particles that are
10 times smaller move 1000 times faster, by Eq.~(\ref{taub}). In
practice, often one can choose to study larger colloidal particles
\shortcite{weeks00} or use careful data analysis techniques to learn
information about smaller-sized particles that aren't directly
imaged \shortcite{kegel04}.  Additionally, when studying dense colloidal
samples, the time scales increase simply because of the glassy
dynamics.

Two main methods of microscopy have been used to study dense
colloidal samples:  conventional optical microscopy and confocal
microscopy.  First, there is the possibility of using a conventional
light microscope technique such as ``brightfield microscopy.''
These techniques typically depend on slight differences between
the index of refraction of the colloidal particles and the solvent
\shortcite{inoue97}.  A limitation is that these differences also
scatter light; each particle acts like a tiny lens.  This ultimately
limits how deeply into a sample one can observe.  (This is the same
phenomenon that makes milk appear white, even though composed of
transparent components; the different components all have different
indices of refraction.  Snow is white for a similar reason, due
to the contrast in index of refraction between the ice crystals
and air.)  A further limitation of conventional optical microscopy
is that the images are limited to a plane, although this can be
fine for quasi-two-dimensional samples \shortcite{marcus99}.

A related conventional technique is fluorescence microscopy.
Here, the particles can be precisely index-matched with the
solvent.  However, the particles also contain a fluorescent dye.
In fluorescence microscopy, the particles are illuminated with
short-wavelength light.  The dye molecules absorb this light,
and radiate slightly longer wavelength (lower energy) light, which
is imaged by the camera.  Special filters and mirrors are used to
direct the light appropriately from the light source to the sample,
and from the sample to the camera.  While this method avoids the
problem of light scattering off of different parts of the sample,
in dense samples it can still be a problem that too much of the
sample fluoresces at the same time, thus giving a large
background illumination.  Trying to observe bright particles on a
bright background thus limits fluorescence microscopy of dense
samples.  One way to overcome this is to only dye a few tracer
particles.

Fluorescence microscopy has one significant limitation:
photobleaching.  After dye molecules absorb the excitation light,
but before they emit light, they can chemically react with oxygen
present in the sample to form a non-fluorescent molecule.  This only
happens when they are excited, so photobleaching happens in direct
proportion to the illumination light.  Photobleaching manifests
itself as the image becoming gradually darker.  This can be a useful
technique for studying local diffusion in samples, a technique known
as ``fluorescent recovery after photobleaching'' \shortcite{axelrod76}.
Intense light is used to photobleach a region of the sample,
and then low-intensity light is used to monitor the recovery of
fluorescence as non-bleached particles diffuse back into the region.
With this method, the diffusivity of the particles can be measured,
which has been used to study the behavior of colloidal glasses
\shortcite{kegel04}.

An extension of fluorescence microscopy is confocal microscopy,
sometimes termed laser scanning optical microscopy.  Here, a
laser is used to excite fluorescence in dye added to a sample.
Typically, the laser beam is reflected off two scanning mirrors
that raster the beam in the $x$ and $y$ directions on the sample.
Any resulting fluorescent light is sent back through the microscope,
and becomes descanned by the same mirrors.  A mirror directs the
fluorescent light onto a detector, usually a photomultiplier tube.

One additional modification is necessary to make a confocal
microscope:  before reaching the detector, the fluorescent light
is focused onto a screen with a pinhole.  All of the light from
the focal point of the microscope passes through the pinhole,
while any out of focus fluorescent light is blocked by this
screen.  This spatial filtering technique blocks out the
background fluorescence light, allowing the particles to be
viewed as bright objects on a dim background.

\begin{figure}[tbp]
\begin{center}
\includegraphics[width=5.5cm]{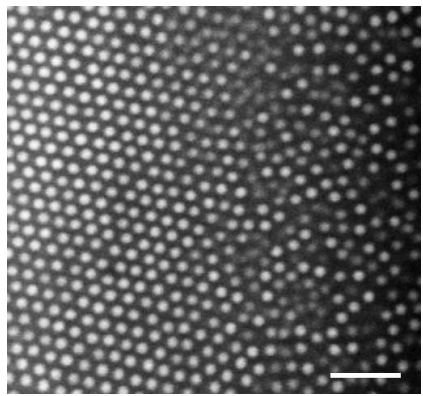}
\end{center}
\caption{ \label{confocalpic} Image of 2 micron diameter colloidal
particles taken with a confocal microscope.  The scale bar is 10
microns.  The sample is in coexistence between a colloidal crystal
and a liquid, see Ref.~\protect\shortcite{hernandez09} for details.  Taken by
Jessica Hern\'{a}ndez-Guzm\'{a}n and Eric R.~Weeks}
\end{figure}

This ability to reject out-of-focus fluorescent light directly
results in the main strength of confocal microscopy, the ability
to take three-dimensional pictures of samples.  By rejecting
out-of-focus light, a crisp two-dimensional image can be obtained,
as shown in Figure \ref{confocalpic}.  The sample (or objective
lens) can be moved so as to focus at a different height $z$ within
the sample, and a new 2D image obtained.  By collecting a stack
of 2D images at different heights $z$, a 3D image is built up.
The time to scan one 2D image can range from 10 ms to several
seconds, depending on the details of the confocal microscope and
the desired image size and quality.  The time to scan a 3D image
depends on the 2D scan speed and the desired number of pixels in
the $z$-direction; reasonable 3D images can be
acquired in 2 - 20~s depending on the microscope
\shortcite{nugent07prl,weeks00}. Finally, we mention Coherent
anti-Stokes Raman scattering (CARS) microscopy \shortcite{kaufman06}, a
technique that allows one to image in 3D colloidal samples
with spatial and temporal resolutions comparable to those of
confocal microscopy.

Once a series of images has been acquired (2D or 3D), the next step
is typically tracking the individual particles within the images.
Within each image, a computer can determine the positions of all of
the particles.  If the particles do not move large distances between
subsequent images, then they can be easily tracked \shortcite{crocker96}.
Specifically, they need to move less between images than their
typical inter-particle spacing.  With confocal microscopy, particles
can be tracked in three dimensions \shortcite{besseling09,dinsmore01}.
This then is the same type of data that is analyzed from computer
simulations.  Simulations, of course, have many advantages,
including a tunability of particle interaction, the ability to
study either Brownian or ballistic dynamics, and more precise
and instantaneous control over parameters such as temperature and
pressure.  The experiments have an advantage that typically the
boundaries are far away:  while perhaps a few thousand particles
might be viewed, they are embedded in a much larger sample with
millions of particles.

\subsection{\label{sec:DLS} Dynamic light scattering}
Dynamic light scattering (DLS)~\shortcite{Berne1976}, also termed photon
correlation spectroscopy, probes the temporal fluctuations of the
refractive index of a sample. In colloidal systems, scattering
arises from a mismatch between the index of refraction of particles
and that of the solvent, so that DLS probes particle density
fluctuations. Experimentally, one measures $g_2(\tau)-1$, the time
autocorrelation function of the temporal fluctuations of the
intensity scattered at a wave vector $q = 4\pi /\lambda
\sin(\theta/2)$. Here, $\theta$ is the scattering angle and
$\lambda$ is the wavelength in the solvent of the incoming light,
usually a laser beam. Under single scattering conditions, the
intensity autocorrelation function is directly related to the
intermediate scattering function $f(q,\tau)$ (ISF, sometimes also
referred to as the dynamic structure factor) :
\begin{eqnarray}
\label{eq:f} f(q,\tau) \equiv \left < N^{-1}\sum_{j,k} \exp\left\{-i
\mathbf{q}\cdot[\mathbf{r}_j(t+\tau)-\mathbf{r}_k(t)]\right\} \right > \nonumber \\
 = \sqrt{\beta^{-1}[g_2(\tau)-1]} =
\sqrt{\beta^{-1}\left[\frac{<I(t+\tau)I(t)>_t}{<I(t)>_t^2}-1\right]}\,
,
\end{eqnarray}
where $\beta \le 1$ depends on the collection optics, $I(t)$ is the
time varying scattered intensity, $\mathbf{r}_j$ the position of the
$j$-th particle, and $N$ the number of particles in the scattering
volume. Note that $f(q,\tau)$ decays significantly when $\Delta
\mathbf{r}(\tau) = \mathbf{r}(t+\tau)-\mathbf{r}(t)$ is of the order
of $q^{-1}$: depending on the choice of $\theta$, DLS probes motion
on length scales ranging from tens of nm to tens of $\mu$m. Another
important point to be noticed is that DLS usually probes collective
motion, since the sum in Eq.~(\ref{eq:f}) extends over all pairs of
particles. However, there are ways to measure the self part of the
ISF by making the contribution of the $j \ne k$ terms vanish from
Eq.~(\ref{eq:f}). This may be accomplished by choosing the
scattering vector $q$ in such a way that $S(q)=1$, where $S(q)$ is
the static structure
factor~\shortcite{PuseyJChemPhys1982,PuseyJPhysA1978}. Alternatively, one
can use optically polydisperse suspensions, as in
Ref.~\shortcite{vanmegen89}, where silica and PMMA particles of nearly
the same radius but different refractive index were mixed. When
matching the average refractive index of the colloids, the measured
ISF contains only contributions from the self terms, $i=k$. Note
that optically composite particles that are polydisperse in size are
usually also optically polydisperse. This is the case, e.g., of PMMA
colloids stabilized by polymer layer with a refractive index
different from that of the core, when the core is polydisperse in
size.

The nature of the averages indicated by the brackets in
Eq.~(\ref{eq:f}) is an important issue. In the definition of the
ISF, the average is over an ensemble of statistically equivalent
particle configurations, while operationally $g_2$ is averaged over
time. Therefore, ergodicity is required for Eq.~(\ref{eq:f}) to
hold. Additionally, in order to reduce noise to an acceptable level,
$g_2$ has to be averaged over at least $10^3-10^4$ $\taua$, with
$\taua$ the relaxation time of the ISF. These requirements are often
impossible to meet for supercooled or glassy colloidal systems,
where $\taua$ can be as large as hundreds of thousands of seconds.
To overcome these difficulties, various schemes have been proposed,
among which the most popular is probably the ``multispeckle''
method~\shortcite{WongRSI1993,BartschJChemPhys1997}. In a multispeckle
experiment, the phototube or avalanche photodiode used in regular
DLS is replaced by a multielement detector, typically a CCD or CMOS
camera sensor. The collection optics is chosen in such a way that
each pixel of the detector corresponds to a different
speckle~\shortcite{Goodman2007}, i.e. to a slightly different scattering
vector. Because distinct speckles carry statistically independent
information, the time average can be replaced in part by an average
over pixels:
\begin{eqnarray}
\label{eq:g2} g_2(\tau)-1 = \left <
\frac{<I_p(t+\tau)I_p(t)>_p}{<I_p(t+\tau)>_p<I_p(t)>_p}-1\right>_t\,
.
\end{eqnarray}
Here $I_p$ indicates the intensity measured by the $p-th$ pixel,
$<\cdot \cdot \cdot >_t$ and $<\cdot \cdot \cdot >_p$ denote
averages over time and pixels, respectively. The set of pixels is
chosen in such a way that they correspond to nearly the same
magnitude of the scattering vector $q$. The number of pixels is
typically of order $10^4-10^5$, so that time averaging only needs to
extend over a few $\taua$. This approach allows very slow and
non-stationary dynamics to be probed effectively.

Although many DLS experiments have been carried on a variety of
glassy colloidal systems, the single-scattering conditions required
by this technique are probably more the exception than the rule. For
mildly turbid suspensions, smart detection schemes, most of which
were pioneered by K. Sch\"{a}tzel~\shortcite{schatzel91}, allow the
rejection of multiply scattered photons, thereby efficiently
suppressing artifacts due to multiple scattering. Popular
implementations of this concept include the so-called two-color and
3-D apparatuses (see Ref.~\shortcite{pusey99} for a review). For very
turbid samples, where photons are scattered a large number of times
before leaving the sample and the contribution of single scattering
is negligible, an alternative formalism has been developed, termed
Diffusing Wave Spectroscopy (DWS)~\shortcite{DWSGeneral}. In a DWS
experiment the intensity correlation function $g_2-1$ is related to
the mean squared displacement, $<\Delta r^2(\tau)>$, rather than to
the ISF, as in DLS. Another important difference is the probed
length scale, which in DWS typically covers the range 0.1 - 100 nm,
much smaller than in DLS. Finally, DWS experiments typically probe
the self motion of the particles, rather than their collective
relaxation.

An alternative way to tackle multiple scattering is provided by
X-photon correlation spectroscopy (XPCS). Modern synchrotron sources
deliver X-ray radiation that is coherent enough to perform the same
kind of experiments as with a laser beam in DLS. Because colloidal
systems scatter X-ray much less efficiently than visible light, in
most cases XPCS measurements can be safely performed in the single
scattering regime without adjusting the refractive index of the
solvent. While long-term beam stability is often still an issue,
several XPCS studies on the slow dynamics of colloidal systems have
been published in the last
years~\shortcite{BandyopadhyayPRL2004,ChungPRL2006,RobertEPL2006,TrappePRE2007,WandersmanJPCM2008,HerzigPRE2009,DuriPRL2009b}.

\subsection{\label{sec:cI} Time and space resolved dynamic light scattering}
In a traditional DLS experiment, the detector is placed in the far
field, so that it collects light scattered by a macroscopic region,
typically of volume 1 $\mathrm{mm}^3$ or more. Additionally, the
intensity correlation function $g_2-1$ has to be extensively
averaged over time. Because of these averages over both time and
space, no information can be a priori extracted on dynamical
heterogeneity, e.g. on the spatial and temporal fluctuations of the
dynamics. In recent years, however, novel light scattering methods
have been proposed to overcome these limitations, providing either
spatially averaged but temporally resolved data (time resolved
correlation, TRC~\shortcite{cipelletti03}), or both spatially  and
temporally resolved measurements (photon correlation imaging,
PCI~\shortcite{DuriPRL2009}).

\begin{figure}[t]
\centering
\includegraphics*[width=.78\textwidth]{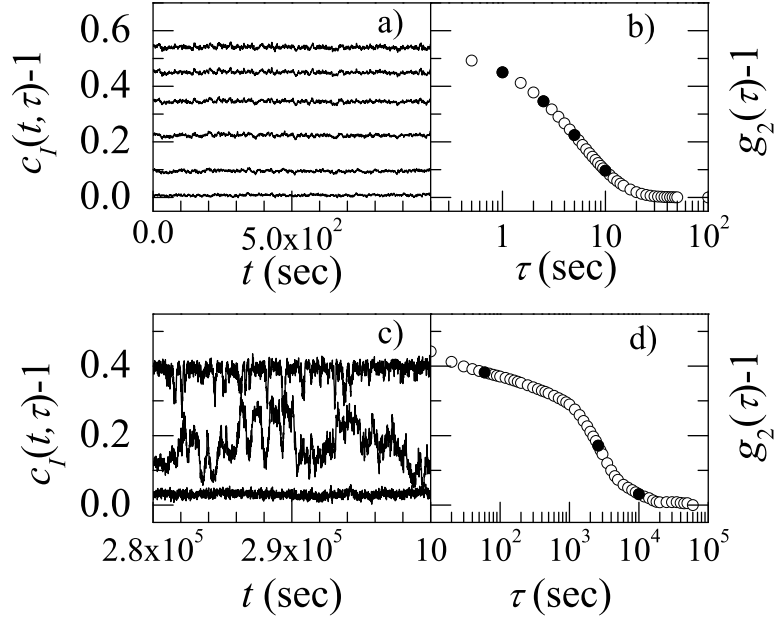}
\caption[]{a) Degree of correlation $c_I(t,\tau)$ for a diluted
suspension of Brownian particles: the dynamics are stationary and
homogeneous, as seen by the very small fluctuations of $c_I$, due
uniquely to measurement noise. From top to bottom, $\tau =
0,1,2.5,5,10$ and 700 sec. b) Intensity correlation function $g_2-1$
obtained by averaging over time the data in a). The solid circles
correspond to the time delays for which $c_I$ is shown in a). c),
d): degree of correlation and intensity correlation function for a
colloidal gel~\protect\shortcite{DuriEPL2006} (see Sec.~\ref{sec:attractive}). In
c), $t=0$ when the gel is formed. Note the large fluctuations of
$c_I$, due to the heterogeneous nature of the dynamics. Individual
events are discernable in the top trace ($\tau = 60$ sec), while the
large fluctuations of the middle trace are due to the superposition
of a fluctuating number of events ($\tau = 2600$ sec). For the
bottom trace, $\tau = 10000$ sec}
\label{fig:cI}       
\end{figure}

In a TRC experiment, one uses a CCD or CMOS detector to calculate a
two-time correlation function $c_I(t,\tau)$ defined by
\begin{eqnarray}
\label{eq:cI} c_I(t,\tau)=
\frac{<I_p(t+\tau)I_p(t)>_p}{<I_p(t+\tau)>_p<I_p(t)>_p}-1\, .
\end{eqnarray}
Note that the usual intensity correlation function $g_2(\tau)-1$
defined in Eq.~(\ref{eq:g2}) is the temporal average of
$c_I(t,\tau)$. Because the detector is typically placed in the far
field, $c_I$ is a temporally resolved but spatially averaged
correlation function. Figure~\ref{fig:cI} shows an example of TRC
data and their relationship to $g_2-1$ for a diluted Brownian
suspension~\shortcite{DuriPRE2005} and for a colloidal
gel~\shortcite{DuriEPL2006}. When plotted as a function of time $t$ for a
fixed time delay $\tau$, the data for the Brownian suspension are
essentially constant. Indeed, for this system the dynamics are
homogeneous and time-translational invariant, so that the evolution
of the system and hence the degree of correlation over a fixed time
lag $\tau$ does not depend on $t$. The small fluctuations around the
mean value are due to the statistical noise of the measurement,
associated with the finite number of pixels over which $c_I$ is
averaged~\shortcite{DuriPRE2005}. For the colloidal gel, by contrast,
$c_I$ exhibits significant temporal fluctuations, indicative of
heterogeneous dynamics. Sudden drops of $c_I$ measured for short
lags, as in the top trace of Fig.\ref{fig:cI}c, are indicative of a
sudden rearrangement event that has led to a loss of correlation
between the intensity patterns recorded at times $t$ and $t$+$\tau$.
As the rearrangement event ceases, the degree of correlation
recovers its typical level. At longer time lags (middle trace in
Fig.\ref{fig:cI}c), $c_I$ has a highly fluctuating behavior, because
several events may occur during the probed lag. By contrast, almost
no fluctuations are observed at very long lags (bottom trace in
Fig.\ref{fig:cI}c), since a large number of events has occurred for
all pairs of images, leading to a full decorrelation.

Various way of characterizing the fluctuations of $c_I$ have been
proposed, including analyzing the probability distribution function
and the moments of its fluctuations, or their temporal
autocorrelation~\shortcite{DuriPRE2005}. Here, we focus on the variance
\begin{eqnarray} \label{eq:chicI} \chi(\tau) = \mathrm{var}(c_I) \equiv \left < \left
[c_I(t,\tau)-<c_I(t,\tau)>_t)\right]^2 \right>_t \, ,
\end{eqnarray}
which is the analogous in light scattering of the dynamical
susceptibility $\chi_4$ discussed in detail in Chapter 2.
Intuitively, one understands that large fluctuations of $c_I$ must
be associated to ``rare'', large events: if the rearrangements were
very localized, many such events would be necessary to significantly
decorrelate the light scattered by a macroscopic sample volume.
Assuming independent events, the resulting spatial average would
yield a smooth $c_I$ trace. More precisely, $\chi$ can be shown to
be proportional to the volume integral of the spatial correlation of
the dynamics, as discussed for $\chi_4$ in Chapter 2. An example of
the scaling of $\chi$ with the size of the events for a coarsening
foam, where events can be unambiguously identified, is discussed
in~\shortcite{MayerPRL2004} (see also Chapter 5). A few differences exist
between $\chi$ in light scattering experiments and $\chi_4$ in
simulations or real space measurements. Contrary to $\chi_4$, $\chi$
is not normalized with the respect to the number $N$ of particles
(compare Eq.~(\ref{eq:chicI}) to eqn~(11) in Chapter 3), since
usually $N$ is not known precisely in light scattering. Accordingly,
typical values reported for $\chi$ are much smaller than those for
$\chi_4$. Moreover, correction methods~\shortcite{DuriPRE2005} to remove
the contribution of the statistical noise to $\chi$ are often used:
using these corrections, one has $\chi = 0$ for homogeneous
dynamics.

\begin{figure}[t]
\centering
\includegraphics*[width=.78\textwidth]{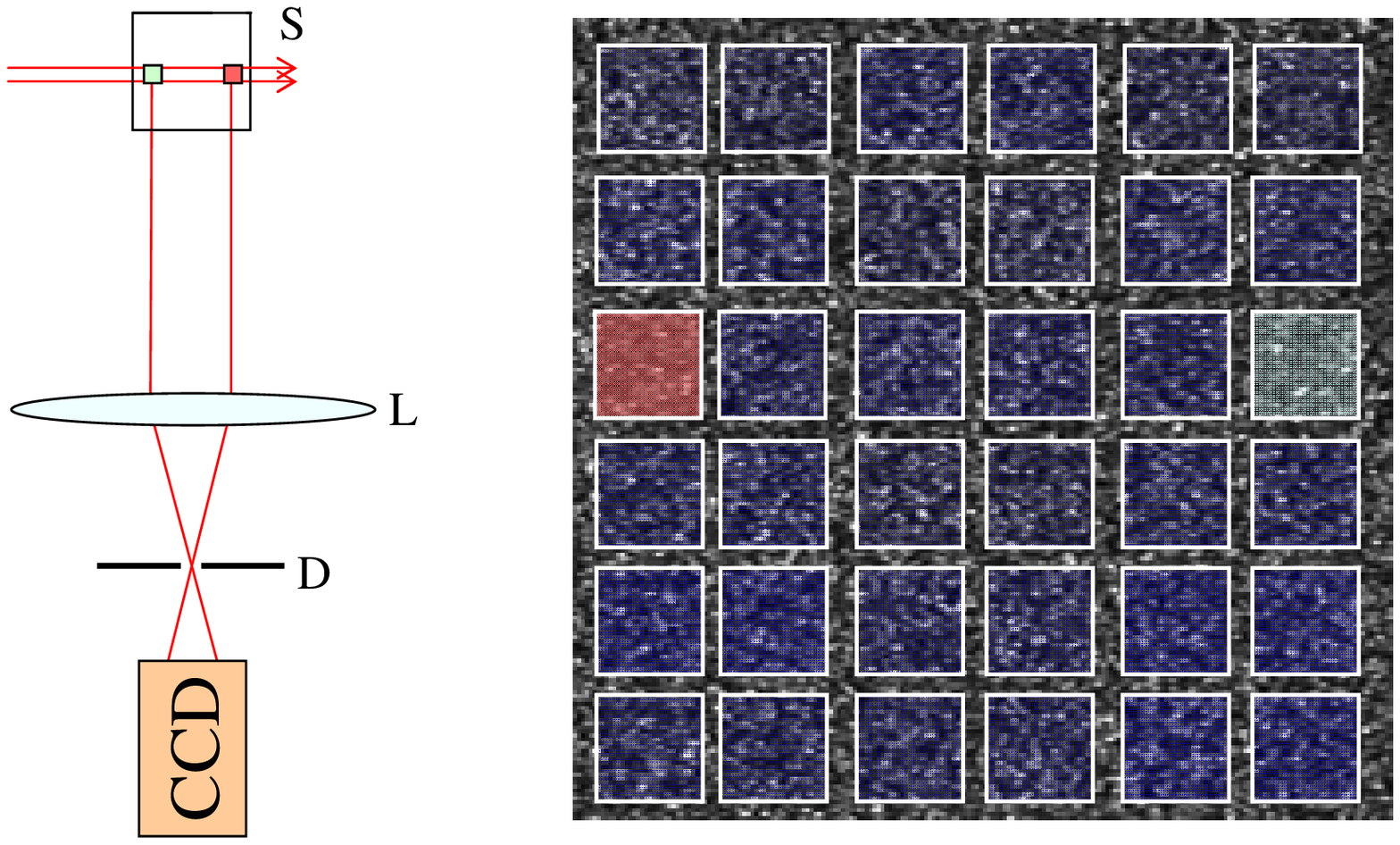}
\caption[]{left: scheme of the Photon Correlation Imaging apparatus
for a scattering angle $\theta= 90^{\circ}$. The lens L makes an
image of the sample S onto the CCD detector. The diaphragm D, placed
in the focal plane of L, selects only light rays scattered at
$\theta \approx 90^{\circ}$. Right: typical CCD image recorded by a
PCI apparatus. The overlaid boxes indicate the grid of ROIs for
which local degrees of correlation,
$c_I(\textbf{\textit{r}},t,\tau)$, are calculated. The size $L$ of
the ROIs has been exaggerated for the sake of clarity; typically,
$L$ is of the order of 10-20 pixels, corresponding to $\sim
20-100~\mu \mathrm{m}$ in the sample}
\label{fig:PCI}       
\end{figure}

While time resolved correlation (TRC) experiments probe temporal
fluctuations of the dynamics, they still lack spatial resolution. In
photon correlation imaging (PCI), by contrast, spatial resolution is
achieved by modifying the collection optics. As shown
in~\shortcite{DuriPRL2009} and sketched in Fig.~\ref{fig:PCI} for $\theta
= 90^{\circ}$, a lens is used to image the scattering volume onto a
CCD or CMOS detector, while a diaphragm limits the range of $q$
vectors accepted by the detector. Under these conditions, each pixel
of the sensor is illuminated by light issued from a small region of
the sample and scattered in a small solid angle associated with the
same, well defined scattering vector. The images are analyzed in the
same way as for TRC, except that they are divided in regions of
interest (ROIs): a local degree of correlation
$c_I(\textbf{\textit{r}},t,\tau)$ is calculated for each ROI, by
averaging the intensity correlation function over a small set of
pixels centered around $\textbf{\textit{r}}$. The spatial
correlation of the dynamics can then be measured by comparing the
temporal evolution of $c_I(\textbf{\textit{r}},t,\tau)$ for ROIs
separated by a distance $\Delta \textbf{\textit{r}}$. More
specifically, we define~\shortcite{DuriPRL2009}
\begin{equation}
g_4(\Delta \textbf{\textit{r}},\tau) = B(\tau)\left < \frac{ \left<
\delta c_I(\textbf{\textit{r}},t,\tau) \delta
c_I(\textbf{\textit{r}}+\Delta \textbf{\textit{r}},t,\tau)\right
>_t}{\sqrt{\mathrm{var}[\delta
c_I(\textbf{\textit{r}},t,\tau)] \mathrm{var}[ \delta
c_I(\textbf{\textit{r}}+\Delta \textbf{\textit{r}},t,\tau)]}} \right
>_{\textbf{\textit{r}}}
\label{eq:G4}
\end{equation}
where $\delta c_I = c_I - <c_I>_t$ are the temporal fluctuations of
the local dynamics and $B(\tau)$ is a normalizing coefficient chosen
so that $g_4(\Delta \textbf{\textit{r}},\tau) \rightarrow 1$ as
$\Delta \textbf{\textit{r}} \rightarrow 0$. This is the analogous,
albeit at a coarse grained level, of the spatial correlation of the
dynamics calculated in numerical and experimental work where
particle trajectories are accessible (see Chapters 2 and 5). In most
cases, the dynamics are isotropic and ${g_4}$ is averaged over all
orientations of $\Delta \textbf{\textit{r}}$. It is important to
distinguish the length scale over which the dynamics are probed from
the spatial resolution with which the local dynamics can be
measured. The former is dictated by the inverse scattering vector.
Depending on the scattering angle, typical values range from a
fraction of $\mu$m up to tens of $\mu$m. The latter is determined by
the size of the ROIs and the magnification with which the sample is
imaged. Typical values are in the range 20-100 $\mu$m. As a final
remark, we note that the differential dynamic microscopy
method~\shortcite{cerbino08} mentioned at the end of
Sec.~\ref{sec:mic_DLS} could be easily adapted to calculate $g_4$.
This would improve the spatial resolution as compared to that of
PCI, thanks to the larger magnification typically used in a
microscope.

\section{Average dynamics and dynamical heterogeneity in the supercooled regime}

\subsection{\label{sec:relax}Structural relaxation time}

\begin{figure}[t]
\centering
\includegraphics*[width=.8\textwidth]{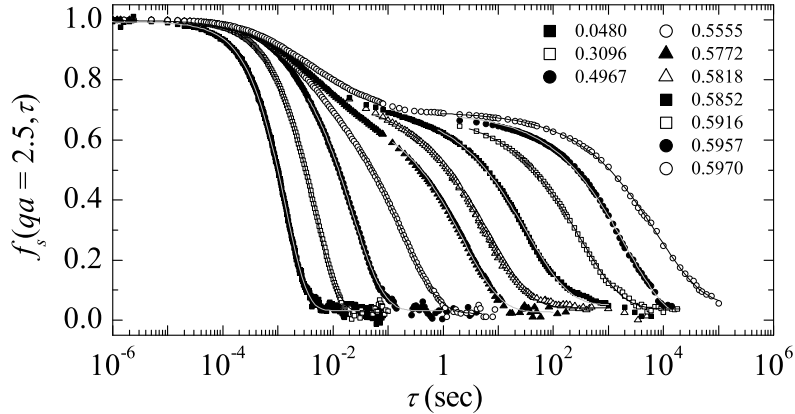}
\caption[]{Intermediate scattering functions for colloidal hard
spheres~\protect\shortcite{brambilla09,brambilla10}. Data are labeled by the
volume fraction $\phi$. The lines are stretched exponential fits to
the final decay of $f_s(q,\tau)$. Adapted from ~\protect\shortcite{brambilla09}
with permission}
\label{fig:fs}       
\end{figure}

The average dynamics of colloidal hard spheres in the supercooled
regime ($\phi < \phi_g$) has been thoroughly studied in a series of
works on PMMA-based
systems~\shortcite{PuseyPRL1987,vanmegen94,vanmegen98,brambilla09,ElMasriJSTAT2009}.
Figure~\ref{fig:fs} shows typical ISFs measured for a variety of
volume fractions at a scattering vector $q = 2.5/a$ ($a = 100$
nm)~\shortcite{brambilla09,brambilla10}, below the first peak of the
static structure factor [we quote here the more precise
determination of $a$ reported in~\shortcite{brambilla10}, slightly
smaller than that in~\shortcite{brambilla09}]. These experiments are
performed close to the best index matching conditions for a PMMA
sample with size polydispersity $\sigma =
12.2\%$~\shortcite{brambilla10}. Under these conditions the sample is
optically polydisperse, as discussed in Sec.~\ref{sec:DLS}; thus,
the self part of the ISF, $f_s$, is probed~\shortcite{ElMasriJSTAT2009}.
At low volume fractions, the decay of $f_s$ is well fitted by a
single exponential, as expected for diluted Brownian particles. As
$\phi$ increases, the ISFs develop a two-step relaxation. The
initial decay depends weakly on $\phi$ and corresponds to the motion
of  a particle in the cage formed by its neighbors. The final decay
corresponds to the relaxation of the cage; its characteristic time,
$\taua$, increases by almost 7 decades in the range of $\phi$
investigated, where all samples equilibrate.

\begin{figure}[t]
\centering
\includegraphics*[width=.8\textwidth]{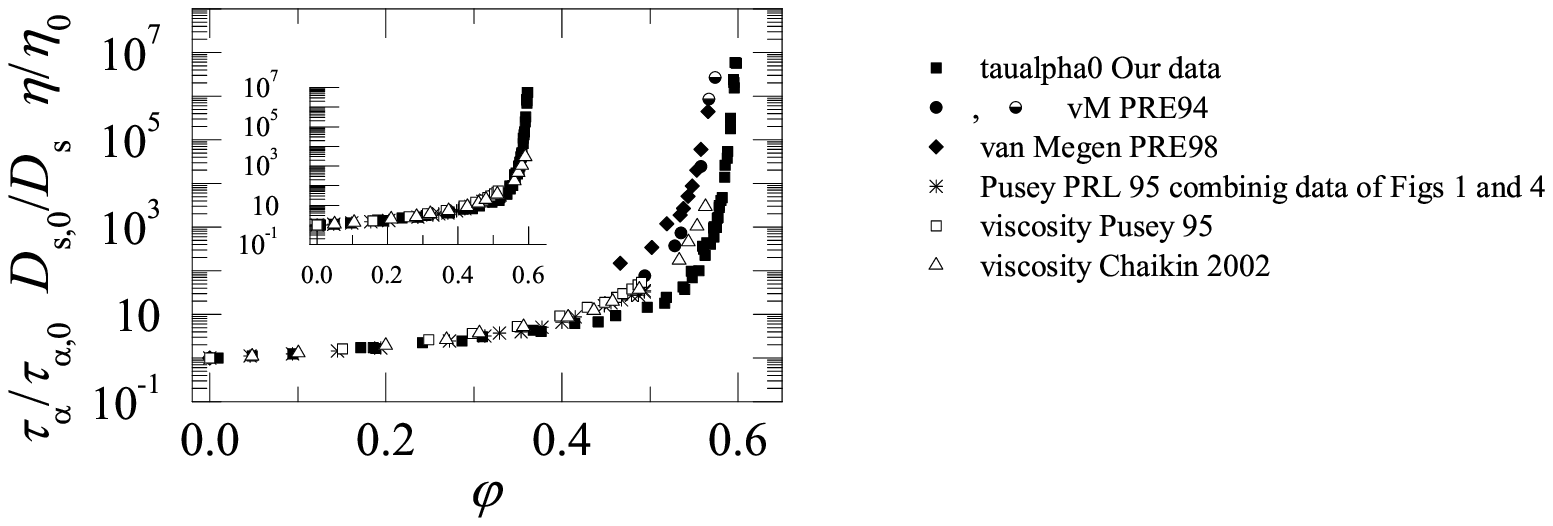}
\caption[]{Main panel: comparison of the volume fraction dependence
of the structural relaxation time, self diffusion coefficient and
low shear viscosity as determined in various works on PMMA colloidal
hard spheres with size polidispersity ranging from $\approx 4\%$ to
12.2\%. All quantities are normalized with respect to their value in
the $\varphi \rightarrow 0$ limit. Solid squares~\protect\shortcite{brambilla09},
solid circles~\protect\shortcite{vanmegen94} and solid diamonds~\protect\shortcite{vanmegen98}
are $\tau_{\alpha}$ data obtained by DLS. Semifilled symbols
indicate data where only a partial decay of $f(q,\tau)$ could be
measured. Stars are self diffusion data obtained by
DLS~\protect\shortcite{SegrePRL1995}. Open squares and open triangles are
viscosity data from Refs.~\protect\shortcite{SegrePRL1995} and~\protect\shortcite{cheng02},
respectively. Inset: same data as a function of scaled volume
fraction, where scaling factors ranging from $1$ to $1.05$  were
chosen so as to superimpose all curves for $\phi \leq 0.2$. [The
data sets from Refs.~\protect\shortcite{vanmegen94,vanmegen98}, for which no
low-$\phi$ data are available, are not included in this plot]. A
reasonably good collapse is obtained also for $\phi > 0.2$,
suggesting that the main source of discrepancy between the various
data sets lies in the uncertainty on the absolute volume fraction
determination}
\label{fig:HSrelaxation}       
\end{figure}

\begin{figure}[t]
\centering
\includegraphics*[width=.7\textwidth]{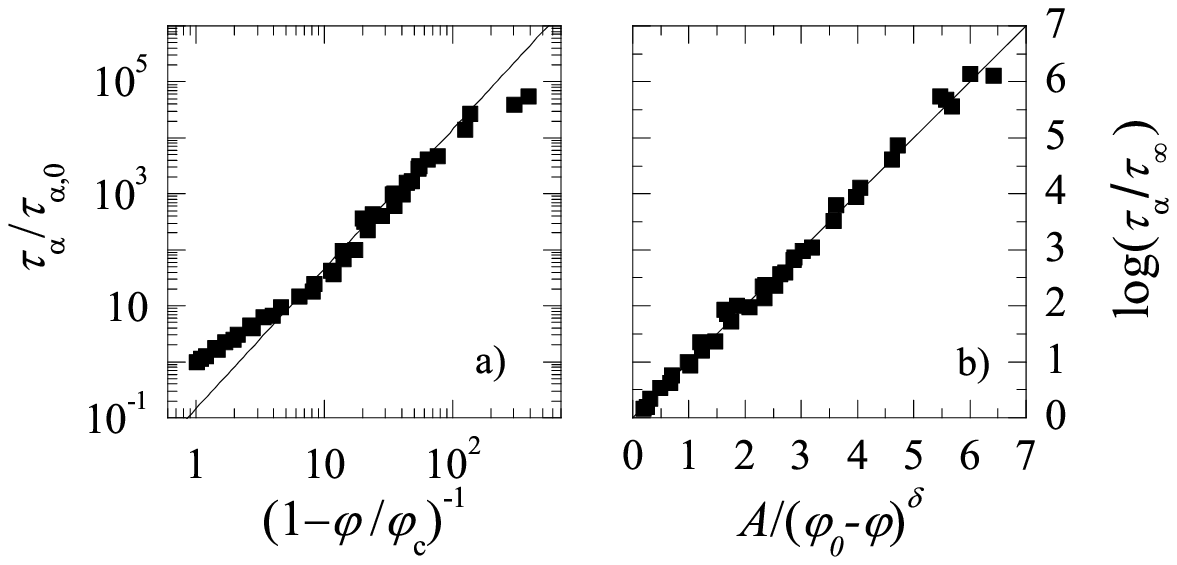}
\caption[]{a): double logarithmic plot of the structural relaxation
time $\taua$ as a function of $(1-\phi/\phi_c)^{-1}$, where the
critical volume fraction $\phi_c$ is obtained from a fit to
Eq.~(\ref{eq:mct}) (adapted from~\shortcite{brambilla09} with
permission). In this representation, the MCT law is a straight line
with slope $\gamma = 2.5$ (solid line). Data at $\phi
> \phi_c$ are not represented in this plot. b): data from Ref.~\protect\shortcite{brambilla09}
for $\phi > 0.41$, plotted using reduced variable such that the
generalized VFT law, Eq.~(\ref{eq:generalizedVFT}) with $\phi_0=
0.637$ and $\delta=2$, corresponds to the straight line shown in the
plot (reproduced from~\protect\shortcite{brambilla09} with permission)}
\label{fig:HSfits}       
\end{figure}

Figure~\ref{fig:HSrelaxation} shows $\taua(\phi)$, as obtained by
fitting the final relaxation of the ISF to a stretched exponential:
\be \label{eq:fsfit} f_s = B \exp[-(\tau/\taua)^\beta] \,,\ee with
$\beta \approx 0.56$ in the glassy regime. In the range $0.517 < \phi
< 0.585$, corresponding to about three decades in relaxation time, the volume fraction dependence of $\taua$ agrees very well
with the critical law predicted by MCT~\shortcite{GotzeJPCM99}: \be
\label{eq:mct}\taua = \tau_0 \left (\frac{\varphi_c}{\varphi_c-
\phi}\right )^{\gamma} \,, \ee with $\gamma = 2.6$ and $\phi_c =
0.59$. This is shown in Fig.~\ref{fig:HSfits}a, where $\taua$ is
plotted against $(1-\phi/\phi_c)^{-1}$. Deviations are observed at
low $\phi$ (as expected, since here MCT does not apply) and, more
importantly, at the highest volume fractions that could be probed,
where $\taua$ grows much slower than expected from MCT. This
suggests that the divergence predicted by MCT is in fact avoided, as
confirmed unambiguously by the fact that equilibrium ISFs could be
measured up to $\phi = 0.598$, above the critical packing fraction
$\varphi_c = 0.59$ obtained from the MCT
fit~\shortcite{brambilla09,ElMasriJSTAT2009}.
Figure~\ref{fig:HSrelaxation} shows also data taken from previous
works on PMMA systems~\shortcite{vanmegen94,vanmegen98}, which exhibit a
similar (apparent) algebraic divergence of $\taua$. Note however
that in these works data at very high $\phi$ could not be obtained,
possibly explaining why the crossover between the MCT regime and the
high $\phi$ ``activated'' regime reported
in~\shortcite{brambilla09,ElMasriJSTAT2009} was not observed previously.

A similar crossover is observed in molecular glass
formers~\shortcite{Donth2001}: the reduced dynamical range accessible in
colloids is responsible for the difficulty in pinpointing it, since
typically $\varphi_c$ is close to $\varphi_g$, the volume fraction
above which equilibrium dynamics become too slow to be
experimentally accessible. The analogy with molecular glass formers suggests that $\taua(\phi)$
may be fitted by a Vogel-Fulcher-Tammann (VFT)-like form: \be
\label{eq:generalizedVFT} \taua  = \tau_{\infty} \exp \left[
\frac{A}{(\phi_0-\phi)^{\delta}} \right] \,, \ee where the exponent
$\delta$ has been introduced for the sake of generality ($\delta =
1$ for the VFT law). Although a very good fit of the data of
Ref.~\shortcite{brambilla09} is obtained for $\delta = 1$, a somehow
better fit is obtained for $\delta = 2$ and $\phi_0 = 0.637$. The
quality of the fit thus obtained can be appreciated in
fig.~\ref{fig:HSfits}b, where the data of Ref.~\shortcite{brambilla09}
are plotted using reduced variables so that
Eq.~(\ref{eq:generalizedVFT}) reduces to a straight line. A crucial
question is whether $\phi_0$ should be identified with
$\phi_{\mathrm{RCP}}$. Experimentally, this question is still open,
due to the difficulties in measuring precisely the volume fraction
at random close packing. Numerical simulations for a binary mixture
of hard spheres~\shortcite{BerthierPRE2009}, however, show that $\phi_0 =
0.641 < \phi_{\mathrm{RCP}} \le 0.664$ supporting the thermodynamic
glass transition scenario and implying that in colloidal hard
spheres the glass and the jamming transition are distinct
phenomena~\shortcite{KrzakalaPRE2007,BerthierPRE2009}. This viewpoint is
still highly debated (see, e.g., \shortcite{XuPRL2009,KamienPRL2007}).

As a final remark on the structural relaxation in colloidal hard
spheres, we note that one may wonder whether size polydispersity may
have a significant impact on the dynamical
behavior~\shortcite{vanmegen10,brambilla10} and in particularly on the
location of the (apparent) MCT divergence. Although no systematic
experiments have been performed to address this
issue~\shortcite{vanmegen01}, computer
simulations~\shortcite{ElMasriJSTAT2009,pusey09,ZaccarelliPRL2009} have
shown that $\varphi_c$ is essentially unaffected by $\sigma$ in the
range $3\%< \sigma < 12\%$ typically explored in experiments.

\subsection{ Viscosity}

Viscosity measurements are an alternative way to probe the average
dynamics of hard sphere suspensions. Several works have been devoted
to the $\phi$ dependence of the viscosity, $\eta$, in the limit of a
vanishingly small applied shear, see e.g. Ref.~\shortcite{cheng02} and
references therein. These measurements are quite delicate, since the
applied stress has to be very small (as low as a fraction of a mPa)
in order to avoid non-linear effects, and because the resulting shear
rate becomes extremely small beyond $\phi \approx 0.5$, limiting
measurements in the deeply supercooled regime. Additionally, the
comparison between results obtained for different samples is
affected by the uncertainties on the absolute volume fraction
discussed above. Cheng \textit{et al.}~\shortcite{cheng02} show that data
from various groups collapse reasonably well on a master curve when
the absolute volume fractions are scaled by a factor up to about
1.03 to account for polydispersity. Figure~\ref{fig:HSrelaxation}
shows viscosity data from Refs.~\shortcite{cheng02,SegrePRL1995} together
with the DLS data discussed in Sec.~\ref{sec:relax}. While there is
some discrepancy between DLS and viscosity data, these differences
are likely to be due, to a great extent, to uncertainties in $\phi$
and to the different methods used in the determination of the
absolute volume fraction. Indeed, data sets for which data points at
low $\phi$ are available can be scaled reasonably well onto a master
curve by correcting $\phi$ using scaling factors between 1 and
1.05, so as to superimpose the growth of the viscosity or the
relaxation time at low volume fraction ($\phi \le 0.2$) (see the
inset of Fig.~\ref{fig:HSrelaxation}). The relationship between
$\eta$ and $\tau_{\alpha}$ is discussed in more detail in
Ref.~\shortcite{SegrePRL1995}, where it is found that the low shear rate
viscosity and the structural relaxation time measured by the
collective ISF at the peak of the structure factor agree remarkably
well up to $\varphi \sim 0.5$, while the relaxation time for the
self part of the ISF is somehow lower.

The nature of the divergence of $\eta$ is still a matter of debate.
In their early work~\shortcite{chaikin96}, Russel, Chaikin and coworkers
reported that $\eta(\phi)$ is fitted well by the Krieger-Dougherty
equation, a critical law of the same form as Eq.~(\ref{eq:mct}) with
$\gamma = 2$, yielding $\phi_c = 0.577$, consistent with the
critical packing fraction obtained from MCT fits of DLS data.
However, in their subsequent analysis of data on a larger range of
viscosity~\shortcite{cheng02}, they report that a VFT-like fit (e.g. of
the form of Eq.~(\ref{eq:generalizedVFT}) with $\delta=1$) yields
better results. They find $\phi_0=0.625$, close to random close
packing, and significantly higher than $\phi_c$. It should however
be recalled that viscosity measurements are feasible only up to
volume fractions lower than $\phi_c$ (e.g. $\phi \leq 0.562$ in
Ref.~\shortcite{cheng02}), making it difficult to draw unambiguous
conclusions on the nature of the divergence of $\eta$, and in
particular on the existence of a divergence at $\phi_c \approx
0.58$.

\subsection{\label{sec:dh_supercooled} Dynamical heterogeneity}

The first microscopy experiment that examined the motion
of supercooled colloidal particles was by Kasper, Bartsch, and
Sillescu in 1998 \shortcite{bartsch98}.  They devised a clever colloidal
system primarily composed of refraction-index-matched particles
made from cross-linked poly-$t$-butylacrylate.  They then added a
small concentration of tracer particles which had non-index-matched
cores of polystyrene coated with shells of poly-$t$-butylacrylate.
Using dark field microscopy, they could observe the motion of the
tracer particles.  They observed that particles exhibited caged
motion, as described above in Sec.~\ref{sec:relax}.  That is,
a particle would diffuse within some local region, trapped in
a cage formed by its neighbors, and then occasionally exhibit a
quicker motion to a new region.  This was useful evidence that
the dynamics are temporally heterogeneous, and the first direct
experimental visualization of caged motion.  Averaging over all
of the tracer particles, they noted that the distribution of
displacements was non-Gaussian, likely linked to the cage
trapping and cage rearrangements.  This experiment
exploited an inherent size polydispersity of about 8\% to prevent
crystallization.  The chief limitations of the experiment were
that the observations were limited to two-dimensional slices of
the three-dimensional sample, and also that only isolated tracer
particles were observed, rather than every particle.

The next published experiment, by Marcus, Schofield, and Rice, had
different tradeoffs \shortcite{marcus99}.  They used a very thin sample
chamber to study a quasi-two-dimensional colloidal suspension; the
spacing between the walls of their sample chamber was approximately
1.2 particle diameters.  Because of the relative ease of studying a
thin sample, they did not need tracer particles, but rather could
follow the motion of every particle within the field of view.  Like
Ref.~\shortcite{bartsch98}, they observed cage trapping and cage
rearrangements, and found a non-Gaussian distribution of
displacements.  Being able to see all of the particles, they also
noted that the cage rearrangement motions were spatially
heterogeneous, with groups of particles exhibiting string-like
motions.  The string-like motions were quite similar to those seen
in simulations~\shortcite{kob97,donati98,hurley96}. Their results clearly
showed a connection between the non-Gaussian behavior and the
spatially heterogeneous dynamics, in that the non-Gaussian
displacements were due to the particles involved in the cage
rearrangements.

The experiment of Kegel and van Blaaderen in 2000 used confocal
microscopy to improve upon the prior experimental limitations
\shortcite{kegel00}.  They observed the motion of core-shell colloidal
particles, in a fully three-dimensional sample (although their
observations were limited to two-dimensional images to maximize the
imaging rate).  Their colloidal samples were well-characterized
as having hard-sphere interactions.  In this experiment, they
again observed string-like regions of high mobility, related to
the non-Gaussian distribution of displacements.  This was the
first experiment to directly visualize spatially heterogeneous
dynamics in a three-dimensional sample, confirming what
simulations had already suggested, that string-like motion is not
an artifact of two dimensional systems
\shortcite{marcus99,kob97,donati98,hurley96,perera99}.

\begin{figure}[tbp]
\begin{center}
\includegraphics[width=8.5cm]{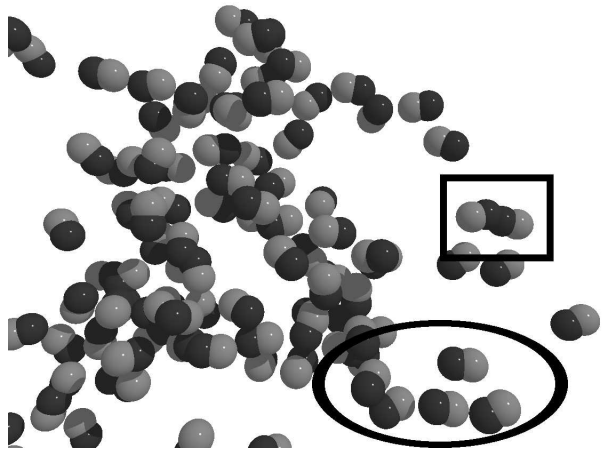}
\end{center}
\caption{ \label{confocalhet} Picture depicting the current
positions of the most mobile colloidal particles (light colored) and
the direction they are moving in (dark colored). The particles are
drawn at 75\% of their correct size, and for clarity only the most
mobile particles are shown.  Many particles move in similar
directions to their neighbors, as shown by those within the oval.
However, some particles also move in opposite directions to their
neighbors, for example closing gaps between them, as highlighted by
the rectangle.  This is a sample with $\phi=0.56$, data taken from
Ref.~\protect\shortcite{weeks00}. The particles have radius $a=1.18$~$\mu$m and
the time scale used to determine the mobility is $\Delta t=1000$~s}
\end{figure}

Shortly after Ref.~\shortcite{kegel00}, Weeks {\it et al.} published a
similar experiment using confocal microscopy to study
three-dimensional colloidal samples~\shortcite{weeks00}. Utilizing a
faster confocal microscope than Ref.~\shortcite{kegel00}, they were able
to observe displacements in three dimensions.  They too saw
string-like motion, although also noted that some particles were
moving in non-string-like ways termed ``mixing'' \shortcite{weeks02}; see
Fig.~\ref{confocalhet}. A limitation of this experiment is that the
particles were later discovered to be slightly charged, rather than
being ideal hard spheres \shortcite{gasser01}.  The three-dimensional
observations enabled the fractal nature of the regions of mobile
particles to be measured as $d_f = 1.9 \pm 0.4$, similar to
simulations \shortcite{donati99}.  The sizes of the mobile regions
increased dramatically as the colloidal glass transition was
approached.

Subsequently, the data of Ref.~\shortcite{weeks00} has been reanalyzed
to highlight other features.  The idea of caging was quantified
in Ref.~\shortcite{weeks02}, finding a result similar to Kasper {\it
et al.} \shortcite{bartsch98}, that caging manifested itself as an
anti-correlation of particle displacements in time.  That is, if
a particle pushed against the ``walls'' of its cage (formed by
its neighbors), then the particle was likely to subsequently be pushed
backwards.  The original work of Ref.~\shortcite{weeks00} highlighted
mobile regions using a slightly arbitrary definition of which
particles were mobile.  Later analysis used spatial correlation
functions to avoid defining a particular subset of particles
as mobile \shortcite{weeks07cor}.  These correlation functions found
length scales for particle mobility which grew by a factor of 2
as the glass transition was approached, finding the largest length
scale for a super-cooled liquid of approximately $8 a$
in terms of the particle radius $a$.

More recent work has taken the study of dynamical heterogeneities in
colloidal suspensions in new directions.  One notable set of
experiments uses superparamagnetic particles and an external
magnetic field to control the glassiness of a sample {\it in situ}
\shortcite{konig05,ebert09,mazoyer09}.  Other experiments study the
colloidal glass transition in confinement \shortcite{nugent07prl}, or
glassy samples as they are sheared
\shortcite{besseling07,schall07,chen10}.

\begin{figure}[t]
\centering
\includegraphics*[width=.9\textwidth]{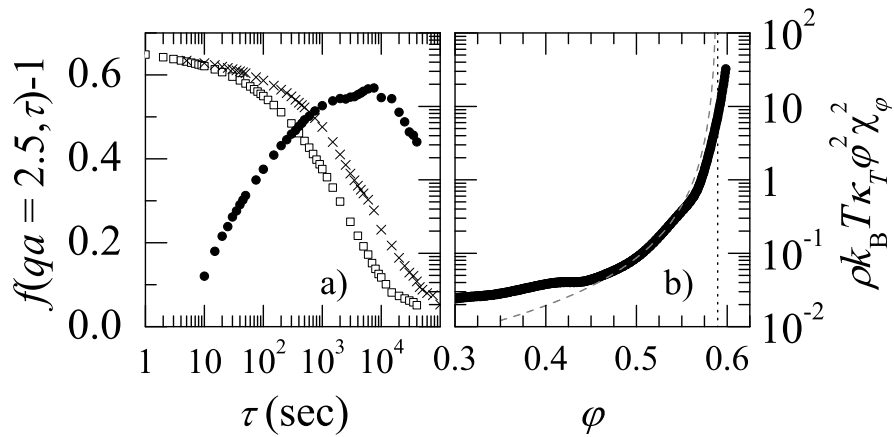}
\caption[]{a), left axis: ISFs at two nearby volume fractions ($\phi
= 0.5953$ and $0.5970$ for the open squares and the crosses,
respectively). The solid circles are $\chi_\phi$ as obtained by
finite difference from the two ISFs (right axis, same scale as in
b)). b): height of the peak of $\chi_4$, $\chi^*$, as a function of
$\varphi$ (adapted from~\protect\shortcite{brambilla09} with permission). Here,
$\chi_4$ has been estimated using the chain rule, as explained in
the text. The dashed line is a MCT fit to the dynamical
susceptibility. The vertical dotted line indicates the location of
the divergence predicted by MCT but avoided by the data}
\label{fig:chi4HS}       
\end{figure}

Dynamical heterogeneity in the equilibrium regime has also been
probed by DLS. Attempts to directly measure $\chi_4$ or $g_4$ using
the time- and space-resolved methods discussed in Sec.~\ref{sec:cI}
have, so far, failed, because these techniques lack the resolution
needed to detect dynamical heterogeneity on the length scale of a
few particles. By contrast, $\chi_4$ can be measured indirectly
using the theory developed in Chapter 3 (see Sec. 3.2.5
therein). For colloidal hard spheres the following relation
holds~\shortcite{BerthierScience2005,BerthierJPhysChem2007a,BerthierJPhysChem2007b}:
\begin{equation}
\chi_4(q,\tau) = \chi_4(q,\tau)|_\phi + \rho k_B T \kappa_T [\phi
\chi_\phi(q,\tau)]^2, \label{eq:chi4equality}
\end{equation}
where $\rho$ is the number density, $\kappa_T$ the isothermal
compressibility (taken from the Carnahan-Starling equation of
state), $\chi_4(q,\tau)|_\phi$ denotes the value taken by
$\chi_4(q,\tau)$ in a system where density is strictly fixed, and
$\chi_\phi(q,\tau) \equiv \partial f_s(q,\tau) / \partial \phi$.
Only the second term in the r.h.s. of Eq.~(\ref{eq:chi4equality})
can be accessed experimentally. Numerical simulations, where both
terms in the r.h.s. of (\ref{eq:chi4equality}) can be
calculated~\shortcite{brambilla09}, show that the first term can be
neglected in the deep supercooled regime: $\chi_4(q,\tau) \approx
\rho k_B T \kappa_T [\phi \chi_\phi(q,\tau)]^2$ when the latter term
is larger than unity. Experimentally, $\chi_\phi$ can be obtained
either by numerical differentiation using two ISFs measured at close
enough volume fractions (see Fig.~\ref{fig:chi4HS}a), or by using
the chain rule in the r.h.s. of Eq.~(\ref{eq:fsfit}):
\begin{equation}\label{eq:chainrule}
\frac{\partial f_s }{\partial \phi }=\frac{\partial f_s}{\partial B}
\frac{\partial B}{\partial \phi}  + \frac{\partial f_s}{\partial
\taua} \frac{\partial \taua}{\partial \phi}  + \frac{\partial
f_s}{\partial \beta} \frac{\partial \beta}{\partial \phi} \,,
\end{equation}
where the partial derivatives with respect to volume fraction of the
coefficients $B$, $\taua$ and $\beta$ defined in
Eq.~(\ref{eq:fsfit}) are obtained by fitting their $\phi$ dependence
by smooth polynomials~\shortcite{Dalle-FerrierPRE2007,brambilla09}.

Figure~\ref{fig:chi4HS}a shows $\chi_4(q = 2.5a,\tau)$ together with
the ISFs used to obtain it by finite
difference~\shortcite{BerthierScience2005}, for the same system as in
Ref.~\shortcite{brambilla09}. The dynamical susceptibility has the
characteristic peaked shape observed in a variety of glassy systems
(see also Chapters 2 and 5), with the maximum of the fluctuations,
$\chi^* \equiv \chi_4(\tau^*)$, occurring on a time scale $\tau^*$
comparable to the structural relaxation time $\taua$. As discussed
in detail in Chapters 2 and 5, $\chi^*$ is of order
$N_{\mathrm{corr}}$, the number of particles that undergo correlated
rearrangements. Figure~\ref{fig:chi4HS}b shows $\chi^*$, obtained
for the same system but using the chain rule~(\ref{eq:chainrule}),
as a function of $\phi$. The amplitude of dynamical fluctuations
increases with volume fraction, supporting theoretical scenarios
where a growing dynamical length accompanies the divergence of the
relaxation time on approaching a glass transition. The dashed line
shows the predictions of advanced mode coupling theories, $\chi^*
\sim (\phi_c -
\phi)^{-2}$~\shortcite{BerthierJPhysChem2007a,BerthierJPhysChem2007b,BiroliEPL2004}.
As observed for the average dynamics, the data agree with MCT only
over a limited range of volume fractions; in particular, MCT
overpredicts the growth of $\chi^*$ at high $\phi$, where the data
remain finite across the the critical volume fraction $\phi_c$.
Therefore, the analysis of dynamical fluctuations confirms the
crossover from an MCT regime to an activated regime at high volume
fractions inferred from the average dynamics.

\section{Average dynamics and dynamical heterogeneity in
non-equilibrium regimes}

\subsection{The glassy regime}

Thus far we have focused on dynamical heterogeneities in supercooled
colloidal liquids; these systems will crystallize after some time
(if the particles are sufficiently monodisperse). For polydisperse
systems, the dynamics can be stationary (unless the sample has been
recently sheared or stirred), since the sample is in (metastable)
equilibrium. In contrast, glasses (volume fraction $\phi > \phi_g$)
are out of equilibrium, and their properties depend on time, a
phenomenon termed ``aging.'' In particular, consider a low volume
fraction colloidal suspension which is centrifuged to rapidly
increase its volume fraction to the point where it becomes a
colloidal glass at some time $t_w=0$.  The motion of particles
within this sample depends on the time $t_w$ since this formation,
called the waiting time or simply the age of the sample. Initially,
particles can move relatively rapidly, but as the system ages,
particle motion slows \shortcite{vanmegen98}. Particles take longer to
move the same distance that was covered quickly at an earlier age.
Equivalently, the cages formed by a particle's neighbors are more
long-lasting, and given that these samples have a high volume
fraction, particles spend almost all of their time tightly confined
within their cages.

In the earliest microscopy experiments (Ref.~\shortcite{bartsch98}),
there was evidence of slow motion within glassy samples.  While
the mean square displacement of the particles grew extremely
slowly, it did grow.  This provided some evidence that particles
were not completely frozen, but rather moved to new positions as
the sample aged.

More direct observations were obtained via confocal microscopy in
Ref.~\shortcite{weeks00}.  In that experiment, Weeks {\it et al.}
observed that while mobile regions of particles grew larger as $\phi
\rightarrow \phi_g$ (in the super-cooled state), in glassy samples
the mobile regions were quite small.  This implied that spatial
dynamical heterogeneities were less important for colloidal glasses.
This result was revised in 2003 by Courtland and Weeks, who observed
larger clusters of mobile particles in an aging colloidal glass,
again using confocal microscopy \shortcite{courtland03}.  The key
difference was in the data analysis. In a glassy sample, most of the
particle motion is Brownian motion within the cages.  Occasionally
particles have cage rearrangements, but this is hard to distinguish
from the Brownian motion as the distances particles move during
these rearrangements gets very small \shortcite{weeks02,courtland03}.
Courtland and Weeks were able to observe the cage rearrangements by
low-pass filtering the raw trajectories, thus smoothing out the
Brownian motion and making the cage rearrangements clearer. They
found that there was indeed spatially heterogeneous dynamics
comparable to the behavior of super-cooled colloids, but
surprisingly they did not see any dependence on the aging time
$t_w$. Similar results were also seen in a later confocal microscopy
study of a binary colloidal glass \shortcite{lynch08}.

One important consideration for studies of the aging of colloidal
glasses is the protocol for achieving the initial glassy state. In
traditional molecular glasses, a sample is quenched from a liquid
state by rapidly reducing the temperature.  In a colloidal glass,
the analogous quench would be to rapidly increase the volume
fraction, for example by centrifugation.  However, in the
experiments described above \shortcite{courtland03,lynch08}, the samples
were instead prepared at a constant volume fraction, and then
shear-melted by stirring them with an embedded stir bar. This is
sometimes termed ``shear-rejuvenation'' as the stirring will make a
well-aged colloidal glass look like a young colloidal glass, that
is, it resets $t_w=0$.  However, there is evidence that these two
protocols give different glassy states in molecular glassformers
\shortcite{mckenna03}.

Motivated by this, two groups have found ways to quench a
colloidal glass {\it in situ} from a low volume fraction state
to a high volume fraction state.  One method uses an external
magnetic field to control the effective inter-particle attraction
in a two-dimensional colloidal sample composed of superparamagentic
particles \shortcite{assoud09}.  This technique has not been used to study
aging explicitly, but so far has focused on slow crystallization
after a rapid quench \shortcite{assoud09}. Another method uses the sample
temperature to control swelling in hydrogel particles, again in a
quasi-two-dimensional experiment \shortcite{yunker09}.  Experiments
using soft
swellable particles are described in Sec.~\ref{sec:soft}.

Dynamical heterogeneity in glassy colloidal samples have also been
probed by DWS. We recall that this light scattering technique is
sensitive to motion on very small length scales, down to a fraction
of a nm (see Sec.~\ref{sec:DLS}), a highly desirable feature for
glassy systems where particles hardly move.
Reference~\shortcite{BallestaNaturePhysics2008} discusses both the
average dynamics and its temporal fluctuations in very dense
suspensions of relatively large particles ($a \approx 10~
\mu\mathrm{m}$). After initializing the sample by shaking it
vigorously, the dynamics slow down until a pseudo-stationary regime
is attained, where all measurements are performed. This regime does
not correspond to a true equilibrium state, but rather to a regime
where the very local dynamics probed in this experiment do not
evolve significantly on the time scale of the experiments (up to a
few days).
\begin{figure}[t]
\centering
\includegraphics*[width=.7\textwidth]{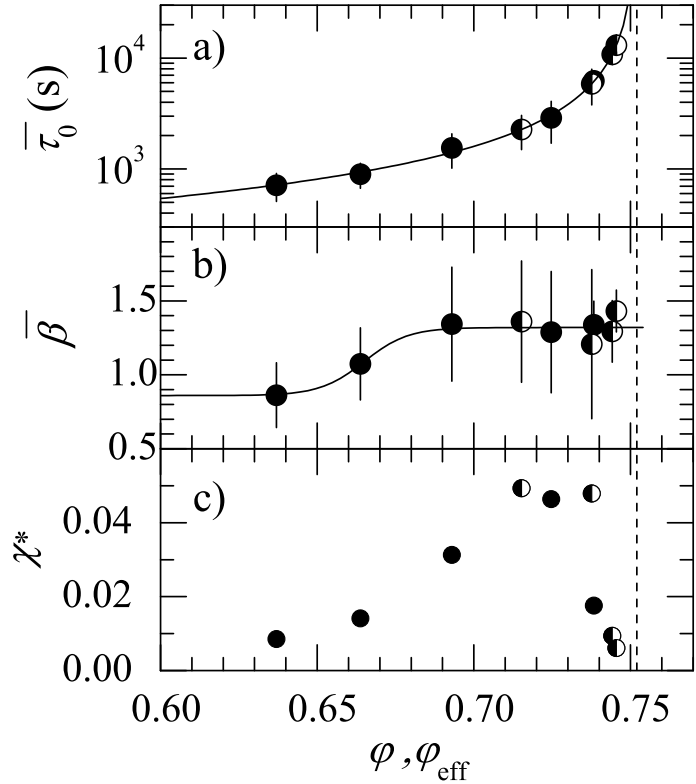}
\caption[]{Volume fraction dependence of the dynamics and its
temporal fluctuations for a concentrated suspension of colloids as
probed by DWS. Solid symbols refer to fresh samples, semi filled
symbols to aged samples that have been rejuvenated mechanically.
See~\protect\shortcite{BallestaNaturePhysics2008} for more details. a) Average
relaxation time, the line is a critical law fit to the data,
$\overline{\tau_0}~\sim (\varphi_\mathrm{max}-\varphi)^{-y}$ with
$\varphi_\mathrm{max} = 0.752$ and $y= 1.5$. Here and in the other
panels the vertical dashed line indicates the location of
$\varphi_\mathrm{max}$. b) Stretching exponent obtained by fitting
the final relaxation of the intensity correlation function to a
stretched exponential. c) Height $\chi^*$ of the peak of the
dynamical susceptibility, $\chi(\tau)$. Note the abrupt drop of
$\chi^*$ on approaching $\varphi_\mathrm{max}$. In a) and b), the
bars indicate the standard deviation of the temporal distribution of
the relaxation time and the stretching exponent, respectively.
Adapted from Ref.~\protect\shortcite{BallestaNaturePhysics2008} with permission}
\label{fig:DWSglass}       
\end{figure}
Figure~\ref{fig:DWSglass} shows the volume fraction dependence of
various parameters characterizing the dynamics in the
pseudo-stationary regime. As shown in a), the average relaxation
time of the intensity correlation function, as obtained from a fit
$g_2(t,\tau)-1 = B\exp[-(\tau/\tau_0(t))^{\beta(t)}]$, increases
smoothly with $\phi$, seemingly diverging at $\phi =
\phi_{\mathrm{max}} = 0.752$, presumably the random close packing
volume fraction for this quite polydisperse system. Panel b) shows
the $\phi$ dependence of the average stretching exponent,
$\overline{\beta}$, which increases from about 0.9 to 1.3 as random
close packing is approached. Panel c) shows the height, $\chi^*$, of
the peak of the dynamic susceptibility $\chi$ defined in
Eq.~(\ref{eq:chicI}). Quite surprisingly, $\chi^*$ is non-monotonic:
it first increases with $\phi$, reaches a maximum, but eventually
drops dramatically close to random close packing.

The interpretation of these data requires special care, since the
usual DWS formalism has been developed for spatially and temporally
homogeneous dynamics. By contrast,
Ref.~\shortcite{BallestaNaturePhysics2008} discusses a general model for
DWS for heterogeneous dynamics, based on an extension of the
formalism originally proposed by Durian \textit{et al.} for DWS
measurements of the spatially localized, temporally intermittent
dynamics of a foam~\shortcite{durian91}. According to the model
of~\shortcite{BallestaNaturePhysics2008}, the non-monotonic behavior of
$\chi^*(\phi)$ results from the competition between two contrasting
effects. On the one hand, the size $\xi$ of the regions that undergo
a rearrangement increases continuously with volume fraction, leading
to enhanced temporal fluctuations as observed in supercooled samples
(see Sec.~\ref{sec:dh_supercooled}). On the other hand, as $\phi$
increases the particle displacement associated with each of these
events is increasingly restrained, due to crowding. Thus, an
increasingly large number of rearrangement events is required to
decorrelate the scattered light at high $\phi$, leading to reduced
fluctuations. While simple simulations of a DWS experiment in a medium undergoing rearrangements described by this scenario
reproduce the experimental data~\shortcite{BallestaNaturePhysics2008}, a
direct measurement of $\xi(\phi)$ would be necessary to confirm it.
This would allow one to better understand analogies and differences
with the experiments on driven grains~\shortcite{LechenaultEPL2008}
mentioned in Chapter 5, where a non-monotonic behavior of
\textit{both} $\chi^*$  and $\xi$ has been reported.

\subsection{Dynamical heterogeneity under shear}

Aging systems are out of equilibrium; another type of
non-equilibrium system is a driven system.  Of particular interest
are samples that are sheared.  The importance of shear can be
quantified by a P{\'e}clet number (see also Sec.~\ref{challenges}).
This is the ratio of the time scale for diffusion to the time scale
for shear-induced motion.  The shear induced time scale is given in
terms of the strain rate as $1/\dot{\gamma}$. Recall that the
Brownian time scale is $\tau_B = a^2/(6D)$ (Eq.~(\ref{taub})).
Typically when considering dense colloidal suspensions, the relevant
diffusion constant is $D_\infty$, the long-time diffusion constant,
which varies with the volume fraction.  The other option would be to
use $D_0$, as given by the Stokes-Einstein-Sutherland formula,
Eq.~(\ref{ses}), which is the diffusion constant in the $\phi
\rightarrow 0$ limit.  If $D_0$ is used, the P{\'e}clet number is
termed the bare P{\'e}clet number, $Pe$, and if $D_\infty$ is used,
it is termed the modified P{\'e}clet number, $Pe^*$. Given that we
wish to understand how particles move and rearrange, this is the
long-time-scale motion ($D_\infty$), and combining the expressions
above we find  $Pe^{*} = a^2 \dot{\gamma} / (6 D_\infty)$. For
$Pe^{*} < 1$, diffusion is the primary influence on particle motion,
and the shear is only a small perturbation.  For $Pe^{*} > 1$, the
shear-induced motion is expected to be more significant. The main
consideration here is that for the shear rate to be significant, an
experiment must have $Pe^{*} > 1$, which is the case for the
experiments described below in this subsection.

Simulations of supercooled fluids found that as $Pe^{*}$ increases,
the system becomes more-liquid like, and dynamically heterogeneous
regions become smaller \shortcite{yamamoto97}, similar to the idea of
shear unjamming a sample \shortcite{liu98}.  As yet, this relationship
between $Pe^{*}$ and dynamical heterogeneity has been untested by
colloidal experiments.

A complementary question is to ask what the nature of a
shear-induced motion is, for a dense amorphous sample.  For example,
sheared crystalline materials respond by the internal motion of
dislocation lines.  Influential simulations by Falk and Langer
found, for amorphous materials, that rearrangements occurred
in localized regions, termed ``shear transformation zones''
\shortcite{falk98}.  In these regions, locally the stress builds
up until it is released by a rapid local rearrangement event.
These behaviors have been directly observed in an experiment by
Schall {\it et al.}, where they used confocal microscopy to
examine a colloidal glass as it was sheared between two parallel
plates at low strains $\sim 4$\% \shortcite{schall07}.
In their experiment, the shear-induced
dynamical heterogeneities had a small spatial extent, although in
some cases they appeared to relax the strain over a large region
even though the farther away particles did not move very far.
Their activation energy was estimated to be $E \sim 16 k_B T$,
where $k_B T$ is the thermal energy.  Given the low strain, it is
possible that these observations are of shear-induced aging
effects, rather than shear flow.

A separate experiment examined the shear-induced motion of colloidal
particles in supercooled colloidal liquids \shortcite{chen10}.  In these
less dense samples, localized rearranging regions were observed
at high strain values.  These samples were sheared between two
parallel plates that moved back and forth with a triangle wave
displacement curve.  Strikingly, the shapes of these rearranging
regions were isotropic on average, and showed no distinction
between the velocity direction, velocity gradient direction,
and the vorticity direction mutually perpendicular to the first
two.  Similar observations were made of a sheared colloidal glass
\shortcite{besseling07}:  in those experiments, little difference was
found in the effective diffusivity in the three directions, once
the affine motion of the average shear profile was subtracted
from the particle displacements.  This study was the first
direction observation of a sheared colloidal glass which directly imaged
shear-induced dynamical heterogeneities
\shortcite{besseling07}. Note that the isotropic nature of
plastic rearrangements observed in experiments is at odds
with recent simulations of a 2D supercooled fluid of soft
particles, where anisotropic structural rearrangements were
observed~\shortcite{furukawa09}.  These observations depended on
careful analysis methods, which have not yet been applied to
experimental data from colloids.

While not directly the same idea as shear-induced dynamical
heterogeneity, it is important to note that many soft glassy systems
-- such as colloids -- exhibit shear-banding
\shortcite{dhont99,dhont03,fielding07,dhont08,fielding09}.  That is, when a large
sample is sheared, sometimes the strain is localized near one of
the boundaries.  Within the ``shear band'' the sample is straining
a significant amount and plastically deforming, while outside the
shear band, the sample is nearly unstrained.  As the stress must be
continuous throughout the sample, this suggests that the material
is acting as if it has two states, a low viscosity state in the
shear band where the sample flows, and an elastic state outside the
shear band without flow.  Shear bands have been noted in colloidal
suspensions \shortcite{vermant03,chen10,besseling07,ballesta08,dhont08},
colloidal gels \shortcite{moller08},
worm-like micelles \shortcite{berret97,olmsted99,cates06},
foams \shortcite{debregeas01,dennin04,hutzler06},
and granular materials \shortcite{losert00,utter08}.  Note that the
experiments discussed in the previous paragraph
(Ref.~\shortcite{besseling07,chen10})
were observations of a homogeneously shearing subregion within
the shear band.

\section{Beyond hard spheres}

\subsection{Soft particles}
\label{sec:soft}
Colloidal particles with soft repulsive potential interactions can
be obtained in several ways. One possibility is to exploit the
softness of an electrostatic or magnetic repulsive potential.
Another possibility is to modify the synthesis of the particles,
e.g. in star polymers or microgel particles, where the degree of
softness is governed by the number of arms and the degree of
crosslinking, respectively. Finally, systems based on the
self-assembly of amphiphilic molecules can form soft spheres, e.g.
with amphiphilic diblock copolymers.

For soft systems, a nominal volume fraction, $\phi\mathrm{_{nom}}$,
is often defined, based on the size of isolated particles and their
number concentration; since particles can be squeezed,
$\phi\mathrm{_{nom}}$ can exceed unity. At low concentration, soft
spheres behave similarly to hard spheres, provided that an effective
size that takes into account the range of the soft repulsion is
used. This is shown, e.g., by viscosity
measurements~\shortcite{RooversMacromolecules1994,BuitenhuisJChemPhys1997,SenffLangmuir1998},
where the increase of the relative viscosity with concentration is
essentially indistinguishable from that of hard spheres. At higher
concentration, however, the behavior deviates
significantly~\shortcite{RooversMacromolecules1994,BuitenhuisJChemPhys1997,SessomsPhilTransA2009,MattssonNature2009}:
the growth of the viscosity or the relaxation time with
$\phi\mathrm{_{nom}}$ is much gentler and samples with
$\phi\mathrm{_{nom}} > \phi_{\mathrm{RCP}}$ may still be fluid,
since particle deformations allow for structural relaxation.

In the past years, aqueous solutions of poly-N-isopropylacrylamide
(PNiPAM) microgels are become one of the most popular soft systems.
Not only can their softness be controlled during the synthesis (by
varying the amount of crosslinking), but it can also be tuned by
varying the temperature, $T$. Under appropriate conditions, a
decrease of a few degrees of $T$ results in a $\sim 20\%$ growth of
the particle radius~\shortcite{SenffLangmuir1998} and in an increased
softness. The influence of the softness of concentrated PNiPAM
particles on their dynamics has been studied in detail by DLS in
Ref.~\shortcite{MattssonNature2009}. Quite remarkably, the authors find
that softness correlates with fragility, defined here as the slope
of $\log\tau_{\alpha}$ $vs$ concentration, $\zeta$, at the glass
transition, in analogy with the definition for molecular glass
formers~\shortcite{Donth2001}, where $1/T$ replaces $\zeta$. Very soft
particles have an Arrhenius-like behavior, $\tau_{\alpha} \sim
\exp(C\zeta)$, harder particles have a larger fragility (steeper
increase of $\tau_{\alpha}(\zeta)$), and hard spheres are the most
fragile system, with the steepest increase of $\tau_{\alpha}$ on
approaching the glass transition. Thus, soft colloids appear as a
promising system for understanding the origin of fragility in
glasses, a long standing problem.

A similar system has been investigated by photon correlation imaging
(PCI) in Ref.~\shortcite{SessomsPhilTransA2009}. Both $\chi$ and $g_4$
have been studied as a function of $\phi\mathrm{_{nom}}$ in
measurements of the collective dynamics at low $q$. The amplitude of
temporal fluctuations of the dynamics is found to increase
monotonically with $\phi\mathrm{_{nom}}$, while the range of spatial
correlations of the dynamics has a non-monotonic behavior, a maximum
being observed at $\phi\mathrm{_{nom}} \approx \phi_{\mathrm{RCP}}$.
Remarkably, spatial correlations of the dynamics here extend over
several mm, in analogy with what was reported for other jammed
materials (see Sec.~\ref{sec:attractive}).

The $T$ dependence of the size of PNiPAM particles provides a
convenient way to quench them rapidly in a glassy state, by
preparing a relatively concentrated ---yet fluid--- sample, which is
then cooled by a few degrees, thereby swelling the particles to a
quenched glassy state in a fraction of a second. This protocol was
used in a quasi-two-dimensional experiment \shortcite{yunker09}, where
observations of aging over nearly six decades in aging time $t_w$
were possible, due to the rapid quench rate. The authors observed
spatial dynamical heterogeneity; particles occasionally underwent
irreversible rearrangements. While the average size of the
rearranging regions remained approximately constant during aging,
similar to the prior study of 3D hard sphere samples
(Ref.~\shortcite{courtland03}), they identified an increase in the size
of a particular class of rearranging regions as the sample aged.
Specifically, the domain size of rearranging particles surrounding
irreversible rearrangements increased during aging. The largest
clusters of rearranging regions involved approximately 100
particles, a size much smaller than the range of dynamical
correlations measured in~\shortcite{SessomsPhilTransA2009}. However, the
technique used to identify rearranging particles limited the
correlation size, making direct quantitative comparisons difficult.
Additionally, they saw a relation between the local structure and
the particle motion, which agrees with some prior work on hard
spheres~\shortcite{cianci06ssc}.

\subsection{Attractive particles}
\label{sec:attractive}

When colloidal particles experience attractive forces, arrested
phases may be obtained also at a volume fraction lower than that
required for glassy dynamics to be observed in hard spheres. The
distinction is often made between colloidal gels (up to $\phi \sim
0.3$) and attractive glasses (over $\phi \sim
0.5$)\shortcite{SandkuhlerCurrOpCollIntSci2004}. Very recent work by
Zaccarelli and Poon further refines the classification of
concentrated attractive systems~\shortcite{zaccarelli09}, based on the
predominance of either caging or bonding. Concentrated, attractive
glasses~\shortcite{EckertPRL2002,PhamScience2002} have a structure
similar to that of repulsive HS systems, while the structure of
diluted gels depends on the strength of the interactions: highly
attractive systems tend to form string-like, fractal structures,
while gel strands are thicker when the interparticle potential well
at contact is close to
$k_{\mathrm{B}}T$~\shortcite{CampbellPRL2005,DibblePRE2006}. The
mechanism leading to gel formation is still intensely debated;
recent work, still controverted, points to the role of an underlying
fluid-fluid phase transition which is arrested once the dense phase
becomes too concentrated~\shortcite{LuNature2008} (see
Ref.~\shortcite{zaccarelli07} for a recent review on colloidal gels).
Experimentally, attractive systems are typically realized either by
screening the Coulomb repulsion in charge-stabilized systems,
thereby exposing the particles to short range, attractive van der
Waals forces~\shortcite{Russell1992}, or by means of the depletion
force~\shortcite{Asakura1958,Vrij1976} induced by adding to the
suspension smaller particles, often polymer
coils~\shortcite{PoonJPCM2002}.

Dynamical heterogeneity in weak gels have been explored mainly by
confocal microscopy~\shortcite{GaoPRL2007,DibblePRE2008} and
simulations~\shortcite{PuertasJChemPhys2004,CharbonneauPRL2007}. The
general picture emerging from these works is that DH is closely
related to structural heterogeneity. The probability distribution
function of the particles displacement (van Hove function) has
typically a non-Gaussian shape, with ``fat'' tails corresponding to
fast particles whose displacement is anomalously
large~\shortcite{GaoPRL2007,DibblePRE2008,ChaudhuriJPCM2008}. These
particles are located at the boundaries of the thick strands
constituting the gel, while the particles buried within the strands
are the least mobile. A detailed analysis of particle mobility as a
function of the number of their neighbors~\shortcite{DibblePRE2008}
confirms this picture. It should be noted that such a structural
origin of DH is in contrast with hard sphere systems, where no clear
connection between DH and structural quantities could be established
so far. The influence of structure on dynamical heterogeneity in
attractive systems has also been highlighted in a series of
simulation papers by A. Coniglio and coworkers (see e.g.
Ref.~\shortcite{FierroJSTAT2008}).

The length scale dependence of dynamical heterogeneity in weak gels
has been explored both numerically~\shortcite{CharbonneauPRL2007} and in
XPCS experiments~\shortcite{TrappePRE2007}. In these works, the peak
$\chi^*$ of the dynamical susceptibility $\chi_4$ has been measured
as a function of scattering vector $q$. $\chi^*$ has a non-monotonic
behavior, the largest dynamical fluctuations being observed on a
length scale of the order of the range of the attractive
interparticle potential. This has to be contrasted to the case of
repulsive systems, where the maximum of $\chi^*$ typically occurs
around the interparticle
distance~\shortcite{CharbonneauPRL2007,DauchotPRL2005}. On a more
technical level, it is worth noting that Ref.~\shortcite{TrappePRE2007}
has demonstrated that modern synchrotron sources and X-ray detectors
are now sufficiently advanced to allow for measurements of dynamical
heterogeneities. The activity in this field is thus growing
rapidly~\shortcite{TrappePRE2007,WandersmanJPCM2008,HerzigPRE2009,DuriPRL2009b,WochnerPNAS2009}
and there is hope that eventually X-ray scattering experiments may
probe dynamical heterogeneity in molecular glass formers and not
only for colloidal systems.

\begin{figure}[t]
\begin{center}
\includegraphics*[width=.9\textwidth]{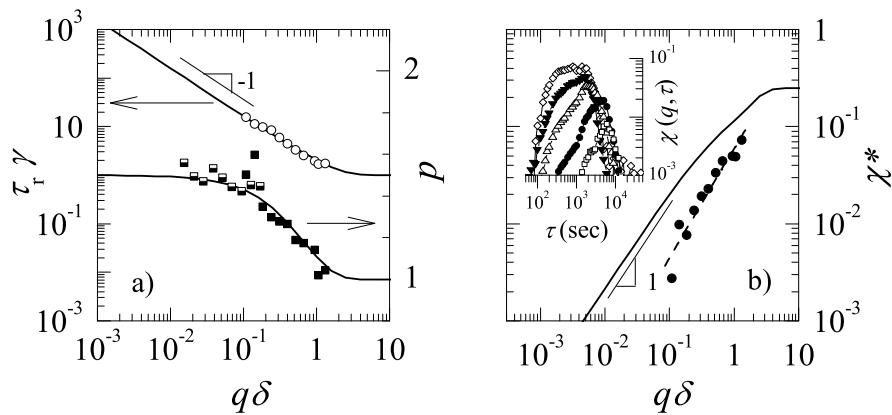}
\end{center}
\caption{
\label{fig:stronggels}       
Average dynamics and dynamical heterogeneity in a strong
gel (adapted from~\protect\shortcite{DuriEPL2006} with permission). {a): $q$
dependence of the relaxation time $\tau_r$ of the intensity
correlation function (left axis, open circles) and of the stretching
exponent (right axis, solid and semifilled squares). The data are
normalized with respect to the parameters of the model (see also the
text): $\delta = 250$ nm is the average particle displacement due to
one single rearrangement event and $\gamma = (960~\mathrm{s})^{-1}$
is the event rate. Note that the motion is, on average,
ballistic-like ($\tau_r \sim q^{-1}$) and that the decay of $g_2-1$
is steeper than exponential (``compressed'' exponential, $p \geq
1$). The lines are the predictions of the model. b) Inset: dynamical
susceptibility $\chi(q,\tau)$ for various $q$. Main plot: $q$
dependence of the height, $\chi^*$, of the peak of $\chi(\tau)$. The
dashed line is a power law fit with an exponent $1.13 \pm 0.11$, the
solid line is the model}
}
\end{figure}

Optical microscopy studies of the dynamics of strong gels are
difficult, due to the restrained and very slow motion of particles
in these systems~\shortcite{DibblePRE2008}. By contrast, scattering
techniques have been successfully applied to characterize the slow
dynamics of tenuous, fractal-like gels made of particles tightly
bound by van der Waals
forces~\shortcite{CipellettiPRL2000,DuriEPL2006,DuriPRL2009}. Time
resolved correlation (TRC, see Sec.~\ref{sec:cI}) experiments show
that the dynamics are due to intermittent rearrangement events where
particles move over relatively small distances, of the order of a
fraction of a $\mu$m~\shortcite{DuriEPL2006}. Quite surprisingly, each of
these events affects a macroscopic portion of the sample: the
spatial correlation of the dynamics measured by photon correlation
imaging (PCI, see Sec.~\ref{sec:cI}) hardly decays over several
millimeters, indicating that spatial correlations of the dynamics
are limited essentially only by the system size~\shortcite{DuriPRL2009}.
Once averaged over both time and space, the intensity correlation
function $g_2(q,\tau)-1$ measured by ``regular'' multispeckle DLS
exhibits a peculiar $q$ dependence of both the relaxation time and
the stretching exponent $p$ obtained by fitting $g_2-1$ to a
stretched exponential, $g_2(q,\tau)-1 \sim
\exp[-(\tau/\tau_r)^p]$~\shortcite{CipellettiPRL2000}. As shown in
Fig.~\ref{fig:stronggels}a, $p$ is larger than one; accordingly, the
relaxation has been termed a ``compressed'' exponential, as opposed
to the as opposed to the stretched exponential relaxations often
observed in glassy systems ($p<1$). Moreover, $\tau_r \sim q^{-1}$,
as opposed to $\tau_r \sim q^{-2}$ as for diffusive motion.

In Refs.~\shortcite{CipellettiPRL2000,BouchaudEPJE2002} these dynamics
were interpreted as ultraslow ballistic motion due to the slow
evolution of a strain field set by internal dipolar stresses. A more
refined model has been proposed in Ref.~\shortcite{DuriEPL2006}, taking
into account the results from time- and space-resolved light
scattering experiments that highlight the discontinuous nature of
the relaxation process. The model is based on the following
assumptions: i) the dynamics are due to individual rearrangement
events that are random in time (Poissonian statistics); ii) each
event affects the whole scattering volume (as indicated by PCI);
iii) the displacement field induced by one single event is that due
to the long-range elastic deformation of the gel under the action of
dipolar stresses (in order to account for $p>1$); iv) on the length
scales probed by the scattering experiments, the displacement due to
successive events occurs along the same direction (i.e. the motion
is, on average, ballistic-like, as implied by the scaling $\tau_r
\sim q^{-1}$). The model contains just two adjustable parameters:
the rate of the events in the scattering volume, $\gamma$, and the
average particle displacement resulting from one single event,
$\delta$. The model captures well the $q$ dependence of both $p$ and
$\tau_r$ as observed in the average dynamics, as shown in
Fig.~\ref{fig:stronggels}a. It also captures correctly the growing
trend for the $q$ dependency of the amplitude of dynamical
heterogeneity, although it overestimates their magnitude by about a
factor of two, as seen in Fig.~\ref{fig:stronggels}b. Physically,
the growth of dynamical heterogeneity with increasing $q$ can be
understood as the result of the competition between the length
scale, $q^{-1}$, over which the dynamics is probed, and the typical
particle displacement, $\delta$, due to a rearrangement event. At
very large $q$, $q \delta \geq 1$, one single event is sufficient to
fully decorrelate $g_2-1$. In this regime, the instantaneous
relaxation time depends on the time between successive events, which
is a fluctuating quantity due to the Poissonian nature of the
events. This yields very large fluctuations of the dynamics. As $q$
decreases, an increasing number of events is required to decorrelate
$g_2-1$, thus leading to smoother dynamics.

Quite intriguingly, the main features of the average dynamics
reported for the strongly attractive gels discussed above have been
also found in a large variety of glassy soft materials (for a
review, see e.g. Ref.\shortcite{CipellettiJPCM2005}). Models inspired by
Ref.~\shortcite{DuriEPL2006} have been used also to describe the dynamics
of nanoparticles embedded in molecular systems approaching the glass
transition~\shortcite{CaronnaPRL2008,GuoPRL2009}. Moreover, photon
correlation imaging
experiments~\shortcite{SessomsPhilTransA2009,maccarrone10} on several
systems (from concentrated soft spheres to repulsive glasses of
charged clays and humidity-sensitive biofilms) suggest that
system-spanning correlations of the dynamics may be a ubiquitous
feature of jammed materials. This is in stark contrast with
supercooled colloidal hard spheres, where spatial correlations of
the dynamics extend over a few particle sizes at most, as discussed
in Sec.~\ref{sec:dh_supercooled}. It is likely that the
predominantly elastic behavior of jammed materials is responsible
for the ultra-long spatial correlations of the dynamics observed in
those systems, since the strain field set by a local rearrangement
can propagate over large distances before being appreciably damped.
Indeed, internal stress relaxation is often invoked as the origin of
theses dynamics~\shortcite{CipellettiPRL2000,BouchaudEPJE2002}, although
a complete understanding of the physical mechanisms underlying this
peculiar yet general relaxation behavior is still lacking.

\section{Perspectives and open problems}

While the average dynamics of glassy colloidal systems has been
intensively studied since the 1980's, experiments on dynamical
heterogeneities started only about twelve years ago, spurred by
advances in microscopy and light scattering methods and stimulated
by numerical works. A (partial) list of what we have learned in the
past years on both the average dynamics and dynamical heterogeneity
of glassy colloids includes:

\begin{list}{$\bullet$}{\setlength{\itemsep}{0ex}}
\item{Microscopy experiments have allowed us to observe directly
what caged motion and cage rearrangements look like.  Data have
been used to quantify the meaning of ``caging" \shortcite{weeks02}.}
\item{Scattering experiments have shed new light on the dynamics of concentrated hard spheres, for which equilibrium dynamics above the ergodic-non ergodic transition predicted by mode coupling theory have been reported~\shortcite{brambilla09}.}
\item{A variety of microscopy and light scattering experiments agree on both the existence and the magnitude of dynamical heterogeneity in several colloidal systems, both 2D~\shortcite{marcus99,konig05,ebert09,mazoyer09} and 3D~\shortcite{bartsch98,kegel00,weeks00}, with both hard~\shortcite{kegel00,weeks00} and soft~\shortcite{konig05,yunker09,SessomsPhilTransA2009} repulsive interactions.  Along with simulations~\shortcite{doliwa98,donati98,glotzer00,yamamoto98}, this provides nice evidence that dynamical heterogeneities are ubiquitous in glassy systems and not artifacts of one particular colloidal system, one particular experimental technique, or one particular simulation method.}
\item{Experiments on colloidal hard spheres have provided some of the first experimental quantitative evidence that the length scale of dynamical heterogeneities increases when approaching a glass transition~\shortcite{weeks00,BerthierScience2005,weeks07cor}.}
\item{Light scattering experiments on deeply jammed attractive or soft repulsive systems~\shortcite{ballesta08,SessomsPhilTransA2009,DuriPRL2009,maccarrone10} have unveiled a richer-than-expected scenario, with ultra-long ranged spatial correlations of the dynamics not observed so far in simulations.}
\end{list}

In spite of these advances, several questions remain open, making dynamical heterogeneity an exciting field of research:
\begin{list}{$\bullet$}{\setlength{\itemsep}{0ex}}

\item{Is there a structural origin of dynamical heterogeneity? While for diluted, attractive systems this has been shown to be the case, for concentrated, repulsive particles a clear answer is still lacking. Progress in this area ¨will likely require the measurement of non-conventional structural quantities, e.g. the ``point-to-set'' correlation function introduced in numerical works~\shortcite{BiroliNaturePhys2008} and discussed in Chapter 2.}
\item{What is the behavior of dynamical heterogeneity in
colloidal glasses and its relationship with aging? While recent
work on soft particles suggests that the slowing down of the
dynamics during aging may be associated with a growth of spatial
correlations of the dynamics~\shortcite{yunker09}, this was not
seen in experiments with harder spheres \shortcite{courtland03};
this question requires further exploration.}
\item{Is the non-monotonic $\phi$ dependence of dynamical fluctuations observed in some systems~\shortcite{BallestaNaturePhysics2008,SessomsPhilTransA2009} a general feature? Although a somehow similar behavior has been reported for granular systems~\shortcite{LechenaultEPL2008}, the explanation proposed for colloids and grains are different: can these contrasting views be reconciled?}
\item{Recent work~\shortcite{DuriPRL2009,SessomsPhilTransA2009,maccarrone10} on jammed soft materials suggests that the relatively short-ranged correlation of the dynamics of supercooled hard spheres may be the exception rather than the rule, since system-size dynamical correlations are observed in those materials. What is the physical origin of these correlations? Why do the numerical simulations not capture these correlations?}
\item{Most work on colloidal systems has been devoted to hard
sphere-like systems. A general understanding of the role of the
interaction potential on the slow dynamics and dynamical
heterogeneity is still lacking.  Recent experiments
show that softer colloids have
strikingly different behaviors than hard colloids
\shortcite{mattsson09}, although these results are not fully
understood.}

\end{list}

Finally, we remark that numerical simulations can nowadays probe a range of relaxation times comparable to that explored by experiments (see e.g.~\shortcite{brambilla09}). On the one hand, this calls for a more rigorous approach to the design of new experiments, since the question of what can be learned from experiments that simulations can not address needs to be asked. On the other hand, this opens the exciting possibility to compare in great detail numerical and experimental results, thereby allowing one to identify the physical mechanisms that are relevant in determining the slow relaxation of glassy colloidal systems.

\section{Acknowledgements}

The work of ERW was supported by NSF Grant No.~CHE-0910707. The work of LC was supported by grants from ACI, ANR, CNRS, CNES, R\'{e}gion Languedoc-Roussillon, Institut Universitaire de France. LC acknowledges many discussions and fruitful collaborations with L. Berthier, G. Biroli, V. Trappe, and D.A. Weitz.
ERW acknowledges helpful discussions and fruitful collaborations
with J. C. Crocker and D. A. Weitz.

\bibliographystyle{OUPnamed_notitle}
\bibliography{eric}
\end{document}